\pdfoutput=1
\documentclass[12pt]{article}
\usepackage[pdftex]{graphicx}
\usepackage{url}
\usepackage{braket}
\usepackage{latexsym}
\usepackage{mathtools}
\usepackage{url}
\usepackage{amsmath}
\usepackage{cite}
\usepackage[bottom]{footmisc}
\DeclareMathOperator{\sgn}{sgn}
\usepackage[utf8]{inputenc}
\begin{document}

\begin{titlepage}
   \begin{center}
       \vspace*{1cm}

        \Huge
       \textbf{Large \(N\) Tensor and SYK Models}
 
       \vspace{0.5cm}
       \vspace{1.5cm}
        
       \large
       \textbf{Jaewon Kim\footnote{Department of Physics, University of California, Berkeley} \\[1cm]{\small Advisor: Igor Klebanov}}
 
       \vfill
        \normalsize
       A thesis presented for the degree of Bachelor of Arts
 
       \vspace{0.8cm}
 
       \includegraphics[width=0.2\textwidth]{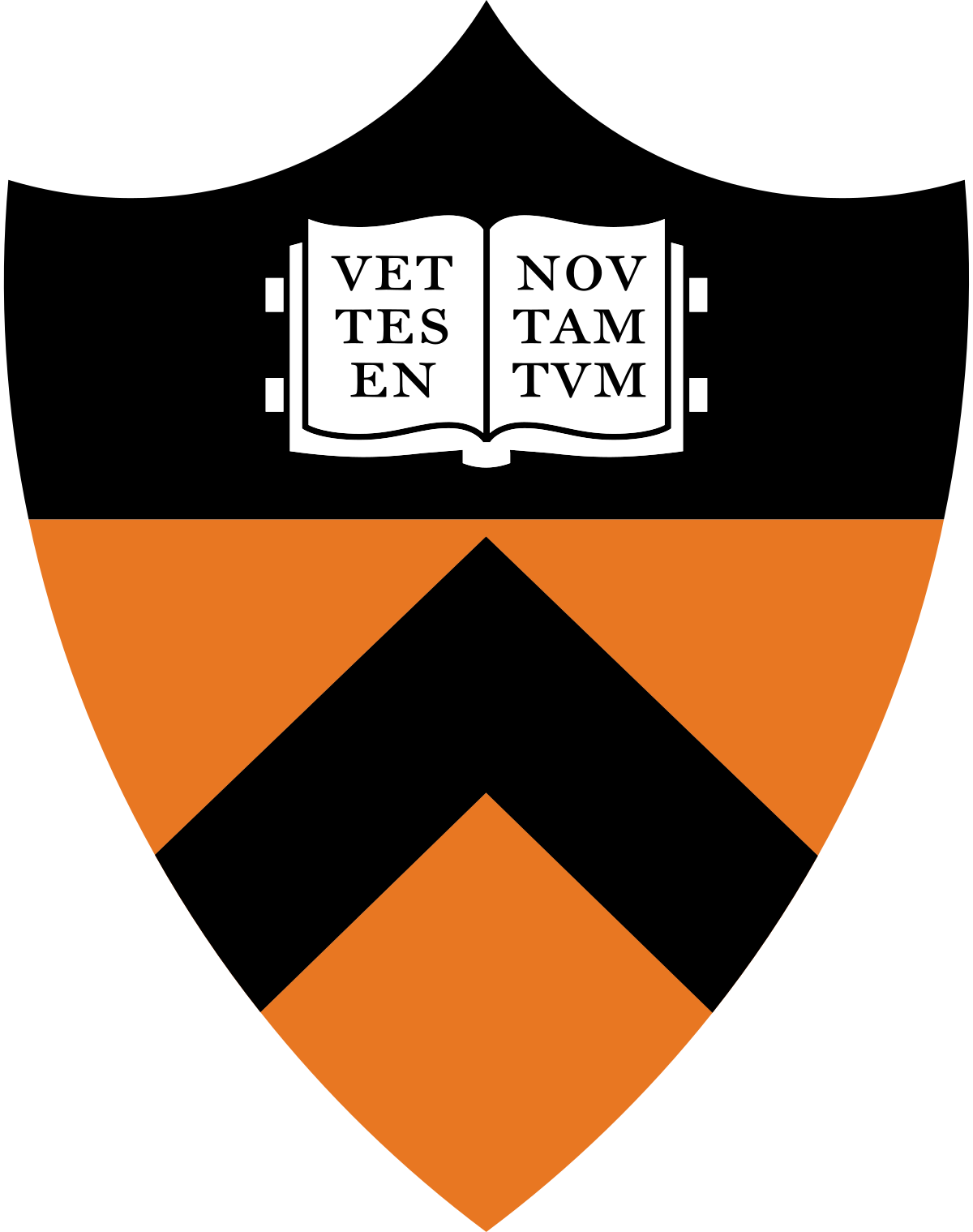}

        \Large
       Department of Physics\\
       Princeton University\\

   \end{center}
\end{titlepage}

\thispagestyle{plain}
\begin{center}
    \Large
    \textbf{Large $N$ Tensor and SYK Models}
 
    \vspace{0.4cm}
    \textbf{Jaewon Kim \\{\small Advisor: Igor Klebanov}}
 
    \vspace{0.9cm}
\end{center}

\begin{abstract}
The SYK model proposed by Sachdev, Ye, and Kitaev consists of Majorana fermions that interact randomly four at a time. The model develops a dense spectrum above the ground state, due to which the model becomes nearly conformal. This suggests that a holographic dual may exist, which makes the SYK model interesting in the study of quantum gravity.
It has been found that the SYK model is similar to large \(N\) tensor models: in both models, only the melonic diagrams survive in the large \(N\) limit.
In this thesis, we explore the large $N$ tensor model with $O(N)^3$ symmetry containing two flavors of Majorana fermions in the fundamental representation. Its quartic Hamiltonian depends on a real parameter $\beta$.
We derive the kernels of the four point functions. With the spectra that we find from the kernels, we calculate the scaling dimensions of several types of conformal primaries. We also find a duality relation between two Hamiltonians of different values of \(\beta\). This is not a perfect duality, because the normalization of energy scales with the transformation. Nevertheless, the ratios of the energies are the same, and the operator dimensions are preserved. In addition, we discover that for \(\beta > 1\) or \(\beta < 0\) the scaling dimensions of one of the conformal primaries become complex, rendering the model unstable.
\end{abstract}

\newpage

\tableofcontents

\newpage
\section{Introduction}
The SYK model is a quantum mechanical model of Majorana fermions which interact with one another randomly. This model, proposed by Kitaev \cite{Kitaev:2015} in his study of black holes and holography, is a variant of the model proposed by Sachdev and Ye \cite{Sachdev:1992fk}. The fact that the SYK model becomes approximately conformal in the infrared, and that it is maximally chaotic, makes it a good candidate for the holographic dual of extremal black holes.

Tensor models, first suggested by Gurau and others \cite{Gurau:2009tw, Gurau:2011xp, Gurau:2011aq, Bonzom:2011zz, Carrozza:2015adg} have been found by Witten \cite{Witten:2016iux} to be similar to the SYK model. Since then, several tensor models have been under extensive study \cite{Klebanov:2016xxf, Gurau:2016lzk, Bulycheva:2017ilt}.

In this thesis, we study a two flavour tensor model in the large \(N\) limit. The model has the same number of degrees of freedom as a complex model, but it only possesses an \(O(N)^3\) symmetry. The Hamiltonian of the model is dependent on a parameter \(\beta\), which controls the strength of the decoupled term of the flavours in the Hamiltonian in relation to the coupled term.

This two flavour tensor model is a generalization of different tensor models that have been studied over the years. By tuning the \(\beta\), one can move between several different SYK-like models that have been studied in various papers. When \(\beta\) is \(\infty\), we arrive at a model which has some similarities with the two-flavor SYK-like model of \cite{Gross:2016kjj} but without the disorder. At \(\beta = -1\), we arrive at the complex tensor model of \cite{Gurau:2016lzk}. When \(\beta\) is 1 or 0, we find the flavourless tensor model studied in \cite{Klebanov:2016xxf}.

In section 1.1 we do a brief review on the SYK model. We introduce the techniques that were utilized to derive the two point functions and the four point functions of the model. In section 1.2, we introduce the various tensor models that have been studied over the years.

In section 2.1, we derive the propagator of the two flavour tensor model. In section 2.2, we calculate the kernels of the four point functions \(\braket{\psi_1^{abc}\psi_1^{abc}\psi_2^{a'b'c'}\psi_2^{a'b'c'}}\), \(\braket{\psi_1^{abc}\psi_2^{abc}\psi_1^{a'b'c'}\psi_2^{a'b'c'}}\), \(\braket{\psi_1^{abc}\psi_2^{abc}\psi_1^{a'b'c'}\psi_2^{a'b'c'}}\) and \(\braket{\psi_1^{abc}\psi_2^{abc}\psi_2^{a'b'c'}\psi_1^{a'b'c'}}\). With the spectra that we obtain from the kernels, we calculate the scaling dimensions of the following conformal primaries: \(\psi_1 \partial_t^{2n+1}\psi_1 \pm \psi_2 \partial_t^{2n+1}\psi_2\), \(\psi_1 \partial_t^{2n}\psi_2 - \psi_2 \partial_t^{2n}\psi_1\) and \(\psi_1 \partial_t^{2n+1}\psi_2 + \psi_2 \partial_t^{2n+1}\psi_1\).

In section 2.3, we go on to find a duality relation between two different values of \(\beta\) through a $\pi/4$ rotation of the Majorana fermions. We go on to check that the spectra obtained from the kernels of the four point functions transform in the same way.

In section 2.4, we look at the instability of this two flavour model at \(\beta > 1\) and \(\beta < 0\): it is found in this section that for \(\beta > 1\) or \(\beta < 0\) the conformal primary operator \(\psi_1 \psi_2 - \psi_2 \psi_1\) has a complex scaling dimension, which renders the model unstable.

\subsection{The SYK Model}
The Hamiltonian of the model in which four fermions interact at a time is given as follows \cite{Kitaev:2015}

\begin{equation}
H = \sum_{iklm}{j_{iklm}\psi_i\psi_k\psi_l\psi_m}
\end{equation}

where \(\braket{j_{iklm}} = 0\), and \(\braket{j_{iklm}^2} =  3!J^2/N^3\)

The more general model in which \(q\) Majorana fermions interact at a time is the following

\begin{equation}
H = i^{\frac{q}{2}}\sum_{1 \leq i_1 < i_2 < \dots < i_q \leq N} j_{i_1, i_2, \dots, i_q}\psi_{i_1}\psi_{i_2} \dots \psi_{i_q}
\end{equation}

Where \(\braket{j_{i_1i_2 \dots i_q}} = 0\), and \(\braket{j_{i_1i_2 \dots i_q}^2} = (q-1)!J^2/N^{q-1}\).

What makes this theory most interesting is that the SYK model is likely to have a holographic dual. The SYK model becomes approximately conformal in the infrared, and there is a reparametrization symmetry which is broken to SL(2,R). This is a charateristic shard by theories of gravity of near extremal blackholes in that they develop a nearly \(AdS_2\) background \cite{Maldacena:2016hyu}.

Furthermore, the chaotic dynamics is also an interesting feature of the SYK model. The growth of the out of time order four point function in this model suggests that inherently, chaotic dynamics are present. This growth of the four point function reveals that the theory is maximally chaotic. This feature matches that of gravity theories. These two features of the SYK model make it an interesting model in the study of quantum gravity, due to the AdS/CFT duality \cite{Maldacena:2016hyu}.

\subsubsection{The Large N Limit of the SYK Model}
The SYK model is solvable in the large \(N\) limit. For example, let us look at the Euclidean propagator \(G(\tau) \equiv \braket{T(\psi(\tau)\psi(0))} = \braket{\psi(\tau)\psi(0)}\theta(\tau) - \braket{\psi(0)\psi(\tau)}\theta(-\tau)\) where \(\theta(\tau)\) is a Heaviside step function. For a free Majorana fermion, the propagator would simply be \(G_{free}(\tau) = \frac{1}{2}\sgn(\tau)\), \(G_{free}(\omega) = -\frac{1}{i\omega}\). Here, we can introduce a finite temperature \(1/\beta\) by letting \(\tau \sim \tau + \beta\) \cite{Maldacena:2016hyu}.

\begin{figure}
  \begin{center}  
    \includegraphics [width=1\textwidth, angle=0.]{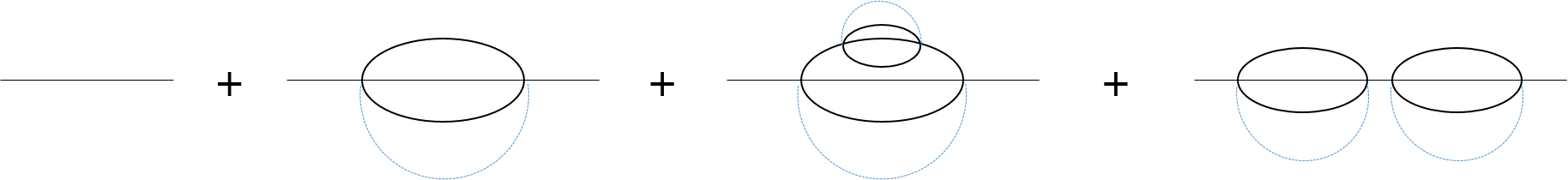}
  \end{center}
  \caption{Perturbative expansion of the SYK Model's two point function. The dotted blue lines denote disorder averaging.}
  \label{fig:SYK}
\end{figure}

By perturbative expansion, one can find the corrections to the two point function caused by the interaction between the fermions. Figure \ref{fig:SYK} illustrates the diagrammatic representation of the perturbative expansion. In doing the perturbative expansion, we perform disorder averaging, and only consider melonic diagrams as illustrated in figure \ref{fig:SYK}. The reason why we do the disorder averaging is simple: As \(\braket{j_{i_1i_2 \dots i_q}} = 0\), diagrams average to 0 unless vertices of the type \(j_{i_1i_2 \dots i_q}\) occur in even numbers.

\begin{figure}
  \begin{center}  
    \includegraphics [width=1\textwidth, angle=0.]{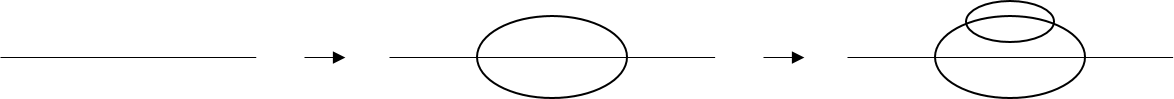}
  \end{center}
  \caption{The iterative procedure of creating melonic diagrams. A propagator as in the leftmost figure is transformed into a two loop diagram as in the middle. Another iterative procedure creates a higher loop-order melonic diagram as illustrated in the rightmost figure \cite{Witten:2016iux}.}
  \label{Melon}
\end{figure}

Moreover, the reason why we only consider melonic diagrams is that the most amount of index contractions occur when the diagrams are melonic. Let us take a moment to define what a melonic diagram is here: As illustrated in figure \ref{Melon}, melonic diagrams are diagrams created by replacing a propagator with a two loop diagram, as illustrated in the middle of the figure \cite{Witten:2016iux}.

\begin{figure}
  \begin{center}  
    \includegraphics [width=1\textwidth, angle=0.]{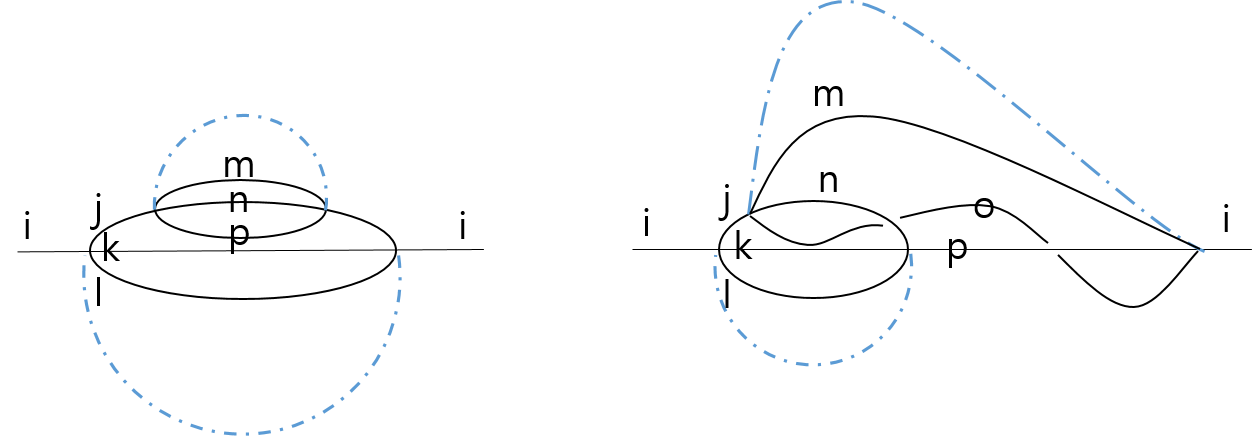}
  \end{center}
  \caption{Left: Melonic Diagram, Right: Non-melonic diagram}
  \label{indexcontraction}
\end{figure}

This thesis does not spend time proving the reason behind melonic dominance in the large \(N\) limit, but we illustrate this point in figure \ref{indexcontraction}. The diagram on the left is melonic, whereas the one on the right is not. The melonic diagram in the left would be summed \(N^6\) times, for there are \(N^6\) ways that we can choose indices j, k, l, m, n, and p. This would result in an amplitude of \(\braket{j^2}^2N^6 = J^4\). On the other hand, for the diagram on the right, we only have \(N^4\) degrees of freedom: due to disorder averaging, i, j, k, l has to equal n, p, k, l, and o, j, n, m, has to equal o, p, i, m. For these relations to hold, n has to equal i, and p, j. As a result we have five indices that we can choose freely, which results in the diagram on the right being summed \(N^5\) times. Hence the amplitude of the diagram would be proportional to \(\braket{j^2}^2N^5 = J^4/N\). Clearly, in the large \(N\) limit the non-melonic term of the right diagram vanishes, while the melonic term of the left diagram stays.

\begin{figure}
  \begin{center}  
    \includegraphics [width=1\textwidth, angle=0.]{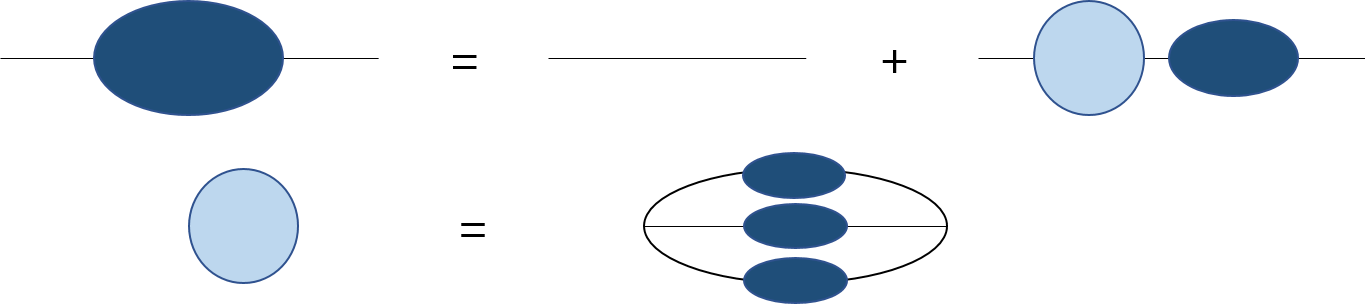}
  \end{center}
  \caption{The navy blue circle diagram denotes the dressed two point function, and the light blue circle diagram denotes the one particle irreducible contributions.}
  \label{twopoint}
\end{figure}

This melonic dominance leads to a simplification in the perturbative expansion of a two point function. This simplification is illustrated in figure \ref{twopoint}.

In the IR limit, we can ignore the left hand term, and this leads us to the following Schwinger-Dyson equation

\begin{equation}
\int{d\tau'} \Sigma(\tau,\tau')G(\tau',\tau'') = -\delta(\tau-\tau''), \qquad\qquad \Sigma(\tau,\tau') = J^2G(\tau,\tau')^3
\end{equation}

In such a limit, the theory has reparametrization invariance. Thus, using the conformal two point function ansatz, one finds

\begin{equation}
G(\tau) = -\Bigg(\frac{1}{4\pi J^2}\Bigg)^{\frac{1}{4}}\frac{\sgn(\tau)}{|\tau|^{1/2}}
\end{equation}

The Appendix provides additional information on the behavior of a two point function in a conformal field theory.

\subsubsection{Four Point Functions of the SYK Model}
\label{SYK4}
Because of the dominance of melonic diagrams in the large \(N\) limit, the SYK model's four point functions become dominated by a particular set of Feynman diagrams called the ladder diagrams. Hence, the four point function of the SYK model can be computed by calculating the ladder diagrams that compose it. In this section, we explain how the four point function \(\braket{\psi_i(t_1)\psi_i(t_2)\psi_j(t_3)\psi_j(t_4)}\) can be computed using this technique, following \cite{Polchinski:2016xgd}.

\begin{figure}[htb]
  \begin{center}  
    \includegraphics [width=0.75\textwidth, angle=0.]{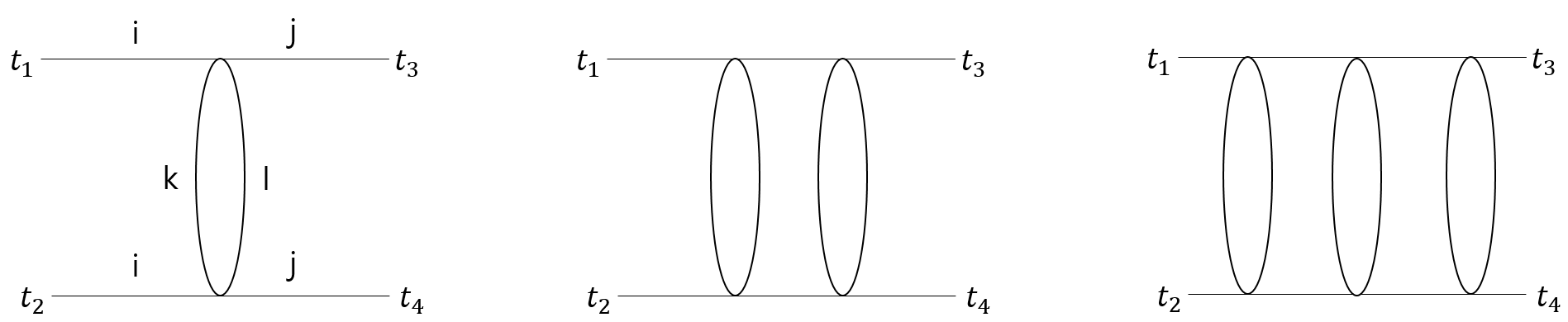}
  \end{center}
  \caption{The four point function consists of ladder diagrams}
  \label{ladders}
\end{figure}

Examples of ladder diagrams are illustrated in figure \ref{ladders}. Notice that the amplitude of the ladder diagrams with a nonzero number of rungs is of order \(1/N\). For example, the leftmost diagram of figure \ref{ladders} has two free indices: one can freely choose k, l. Hence, the amplitude of the diagram is proportional to \(\braket{j^2}N^2 = J^2/N\). In general, all ladder diagrams of the SYK model are of order \(1/N\).

\begin{figure}[htb]
  \begin{center}  
    \includegraphics [width=1\textwidth, angle=0.]{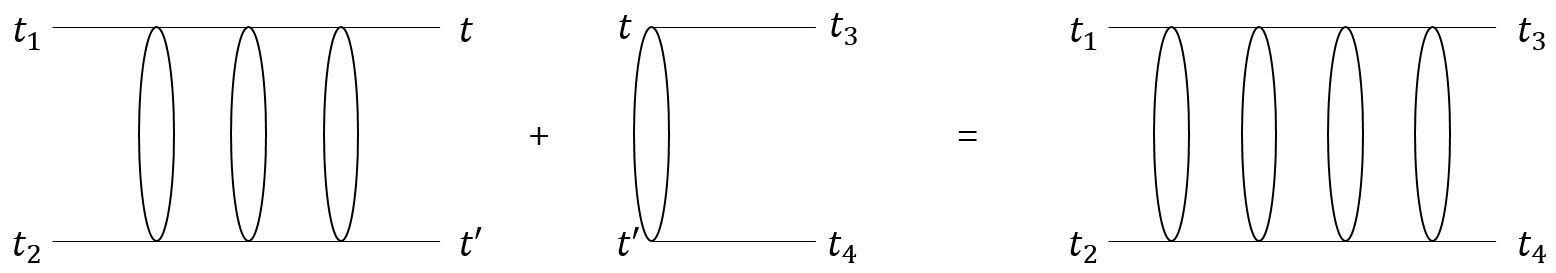}
  \end{center}
  \caption{Adding the kernel (the middle diagram) into a ladder diagram (as on the left) creates a ladder diagram with one more rung (as on the right)}
  \label{ladker}
\end{figure}

The ladder diagrams themselves can be computed by finding the kernels that comprise the diagrams. Kernels are what one needs to add to an \(n\) rung ladder diagram to make it into an \(n+1\) rung ladder diagram. An example of this process is drawn in figure \ref{ladker}.

The kernel for the SYK model is given as the following

\begin{equation}
K(t,t',t_3,t_4) = -3J^2G(t-t')^2G(t-t_3)G(t'-t_4)
\end{equation}

Using the result for $G(\tau)$ that we obtained in the previous section

\begin{equation}
K(t,t',t_3,t_4) = -\frac{3}{4\pi}\frac{\sgn(t-t_3)\sgn(t'-t_4)}{|t-t_3|^{1/2}|t'-t_4|^{1/2}|t-t'|}
\end{equation}

Let us call the \(n\) runged ladder diagram of the four point function \(\braket{\psi_i(t_1)\psi_i(t_2)\psi_j(t_3)\psi_j(t_4)}\) \(\Gamma^n_{SYK}(t_1,t_2,t_3,t_4)\). Then, the following equation holds

\begin{equation}
\begin{split}
\MoveEqLeft
\Gamma^{n+1}_{SYK}(t_1,t_2,t_3,t_4) = \int{dtdt'}\Gamma^n_{SYK}(t_1,t_2,t,t')K(t,t',t_3,t_4)
\end{split}
\end{equation}

Therefore, writing \(O_{SYK}(t_1,t_2,t_3,t_4)\) as the sum of the connected ladder diagrams, it satisfies the following

\begin{equation}
\begin{split}
\MoveEqLeft
O_{SYK}(t_1,t_2,t_3,t_4) = \sum_{n=1}^{\infty}\Gamma^n_{SYK}(t_1,t_2,t_3,t_4) \\&\qquad\qquad\qquad = \Gamma^1_{SYK}(t_1,t_2,t_3,t_4) \\&\qquad\qquad\qquad\qquad\qquad+\int{dtdt'}O_{SYK}(t_1,t_2,t,t')K(t,t',t_3,t_4)
\label{SDF}
\end{split}
\end{equation}

In low energies/high interactions the theory is conformal. Thus, this symmetry can be used to diagonalize the kernel. Define \(v_{SYK}(t_1, t_2)\) as the following

\begin{equation}
v_{SYK}(t_1, t_2) = \frac{\sgn(t_1-t_2)}{|t_1-t_2|^{h}}
\end{equation}

Then, the following holds

\begin{equation}
\int{dtdt'}v_{SYK}(t, t')K(t, t', t_1, t_2) = g(h)v_{SYK}(t_1, t_2)
\end{equation}

where \(g(h) = -\frac{3}{2}\frac{1}{(1-h)\tan(\pi h/2)}\).

Now, using the \(SL(2,R)\) invariance, it is possible to find the complete set of eigenfunctions. They turn out to be

\begin{equation}
v_{\mu \omega}(t_1,t_2) = \frac{\sgn(t_1-t_2)}{\sqrt{4\pi}|t_1-t_2|}e^{-i\omega(t_1+t_2)/2}\Big(J_{\mu}(|\omega(t_1-t_2)/2|)+\frac{\tan(\mu\pi/2) + 1}{\tan(\mu\pi/2) - 1} J_{-\mu}(|\omega(t_1-t_2)/2|) \Big)
\end{equation}

With this complete set of eigenfunctions and equation \ref{SDF}, the four point function \(O_{SYK}\) is found to be the following

\begin{equation}
O_{SYK}(t_1,t_2,t_3,t_4) = \frac{3J}{\sqrt{4\pi}}\int{d\mu d\omega}\frac{v^*_{\mu \omega}(t_1,t_2)v_{\mu \omega}(t_3,t_4)}{1-g(\mu)}
\end{equation}

where \(N_{\mu} = (2\mu)^{-1}\) for \(\mu = 3/2 + 2n\), and \(\mu = (2\mu)^{-1}\sin(\pi\mu)\) for \(\mu = ir\).

Using similar techniques, Gross and Rosenhaus found that any \(N\) point function of the SYK model can be explicitly computed in the large \(N\) limit with the operator product expansion \cite{Gross:2017aos}.

\subsection{Tensor Models}
Just as the SYK model is solvable in the large \(N\) limit, quantum field theories with a large number of fields related by symmetries are greatly simplified in the large \(N\) limit. In a way, this limit acts just like a classical limit would, where in this case the classical limit is achieved in the dual gravitational theory \cite{Gross:2017aos}. 

\begin{figure}
  \begin{center}  
    \includegraphics [width=0.3\textwidth, angle=0.]{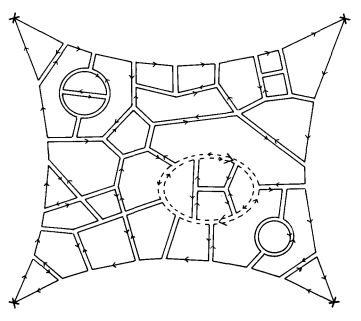}
  \end{center}
  \caption{For the matrix models, planar diagrams as drawn above dominates \cite{'tHooft:1973jz}}.
  \label{planardiagram}
\end{figure}

Vector models were the first to be studied in the large \(N\) limit. The scalar \(O(N)\) vector model had the interaction term \(\frac{1}{4}\lambda\phi_a\phi_a\phi_b\phi_b\), and was easily solved in the large \(N\) limit where \(gN\) is fixed. In this limit tadpole diagrams dominate, and summation over these diagrams is fairly straightforward \cite{Moshe:2003xn}. Then came matrix models, with the interaction term  \(\phi_{ab}\phi_{bc}\phi_{cd}\phi_{da}\). In the large \(N\) limit where \(gN\) is fixed, planar diagrams dominate the perturbative expansion \cite{'tHooft:1973jz}; an example of a planar diagram is given in figure \ref{planardiagram}. This allows for a solution in low dimensional cases, although the theory is not solvable in general \cite{Brezin:1977sv}.

Naturally, what came next were tensor models \cite{Ambjorn:1992iz, Sasakura:1990fs, Gross:1991hx}. Gurau and others found ways to tune the interaction terms of a tensor field so that a tensor model of rank equal to or larger than 3 can be solvable in the large \(N\) limit \cite{Gurau:2009tw, Gurau:2011xp, Gurau:2011aq, Bonzom:2011zz, Carrozza:2015adg}. Subsequently, Witten generalized Gurau's construction of rank 3 tensors from the \(d = 0\) tensor integral case to a \(d\) dimensional QFT.  The interaction term of the Gurau - Witten model is given as follows \cite{Witten:2016iux}

\begin{equation}
V = g\psi_0^{abc}\psi_1^{ade}\psi_2^{fbe}\psi_3^{fdc} + c.c.
\end{equation}

As will be discussed in the next section, the tensor models are dominated by melonic diagrams in the large \(N\) limit, where \(g^2N^3\) is held fixed. This dominance of melonic diagrams at large \(N\) is a characteristic that is shared with the SYK model \cite{Witten:2016iux}: this particular similarity prompted Witten to put forward his \(U(N)^6\) symmetric Gurau-Witten tensor model as a disorderless replication of the SYK-like model.

\subsubsection{The Dominance of Melonic Diagrams at Large \(N\)}
\begin{figure}
  \begin{center}  
    \includegraphics [width=0.3\textwidth, angle=0.]{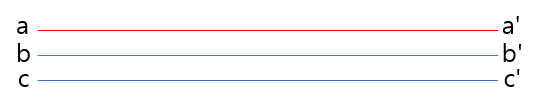}
  \end{center}
  \caption{The propagator \(\braket{\phi_{abc}\phi_{a'b'c'}} = \delta^{aa'}\delta^{bb'}\delta^{cc'}\).}
  \label{tkp}
\end{figure}

\begin{figure}
  \begin{center}  
    \includegraphics [width=1\textwidth, angle=0.]{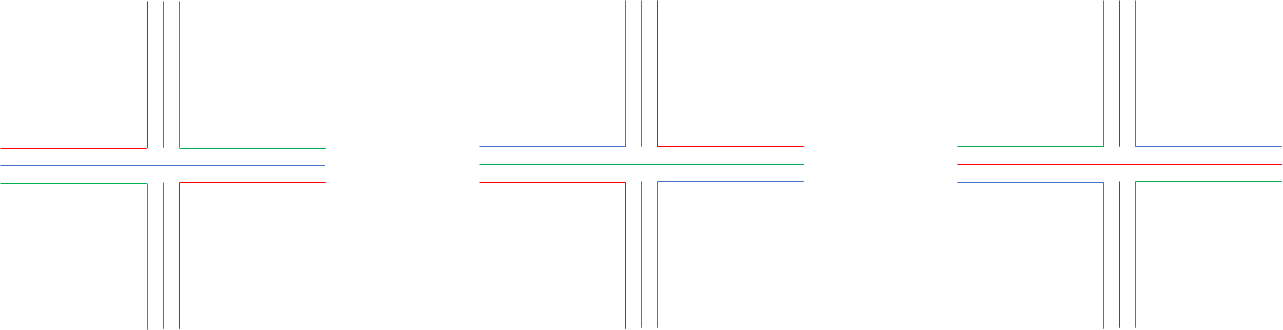}
  \end{center}
  \caption{Vertex in the Klebanov - Tarnopolsky Model.}
  \label{tkv}
\end{figure}

The tensor models that Gurau and Witten proposed exhibit a dominance of melonic diagrams at large \(N\). In this section, we use the model studied by Klebanov and Tarnopolsky as an example. Their model has the interaction term \(\frac{1}{4}g\psi^{a_1b_1c_1}\psi^{a_1b_2c_2}\psi^{a_2b_1c_2}\psi^{a_2b_2c_1}\), and consequently the propagator and vertex functions take the forms that are illustrated in figures \ref{tkp} and \ref{tkv}\cite{Klebanov:2016xxf}.

\begin{figure}
  \begin{center}  
    \includegraphics [width=0.5\textwidth, angle=0.]{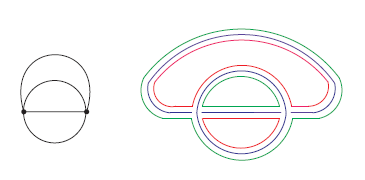}
  \end{center}
  \caption{This diagram has 6 index loops, and hence will be summed up \(N^6\) times \cite{Klebanov:2016xxf}. The number of index loops can be counted by adding together the number of index loops in each subgraph created by a single color.}
  \label{ktexample}
\end{figure}

In perturbative expansions, a diagram with \(\mathcal{N}\) index loops have an amplitude that is summed up \(N^{\mathcal{N}}\) times: For example, the diagram in figure \ref{ktexample} has 6 index loops, and hence gets added \(N^6\) times.

Consequently, a diagram with the most number of index loops at a given number of vertices exhibits the largest amplitude. In the Klebanov - Tarnopolsky model, for a fixed number of vertices, melonic diagrams have the most amount of loops. It is not to difficult to prove this, and the proof is given in \cite{Carrozza:2015adg}. A quick example of the dominance of melonic diagrams is as follows.

\begin{figure}
  \begin{center}
    \includegraphics [width=1\textwidth, angle=0.]{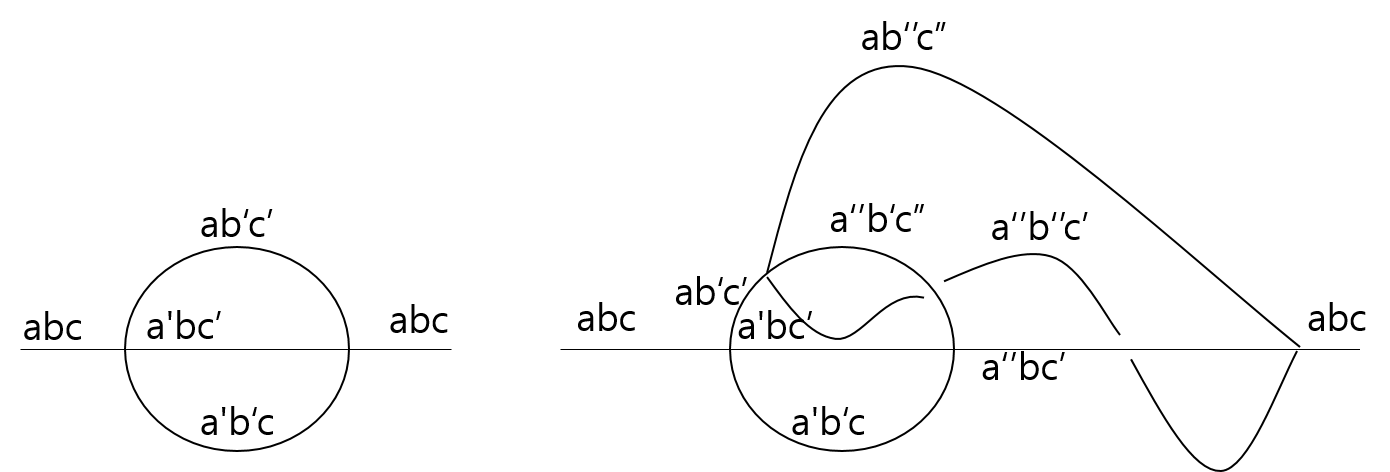}
  \end{center}
  \caption{Diagrams that contribute to \(\braket{\psi_{abc}\psi_{abc}}\). Left is melonic, right is not.}
  \label{tkmelon}
\end{figure}

In the left diagram of figure \ref{tkmelon} there are two insertions of the Hamiltonian. Since the free propagator is a delta function, the vertices must be of the same type. Hence, summing up over \(a\) \(b\) \(c\), and \(a'\) \(b'\) \(c'\), one sums the diagram up \(N^6\) times. This term is therefore expected to have an amplitude proportional to \(g^2N^6\).

On the other hand, in the right diagram of figure \ref{tkmelon}, there are four insertions of the Hamiltonian. Doing a similar analysis we find that at the rightmost vertex, propagators of the type \(\psi_{ab''c''}\), \(\psi_{a''b''c'}\), \(\psi_{a''bc'}\), and \(\psi_{abc}\) meet. Due to the constraint induced by the Hamiltonian, \(c'' = c\). All in all, in summing up over \(a\) \(b\) \(c\), \(a'\) \(b'\) \(c'\), and \(a''\) \(b''\) \(c''\), one adds the diagram \(N^8\) times. Therefore, the term is expected to have an amplitude proportional to \(g^4N^8\).

In the large \(N\) limit, \(g^2N^3\) is kept constant while \(N\) is taken to infinity. Hence, the right diagram of figure \ref{tkmelon} scales as \(O(1/N)\) compared to the left diagram and gets dwarfed by it in the large \(N\) limit. Similarly, all non-melonic diagrams amplitudes go to zero in the large \(N\) limit.

\subsubsection{Flavourless Tensor Model, Complex Bipartite Tensor Models, Bosonic Tensor Models}
This model mentioned in the above section was first explored by Carrozza and Tanasa in \(d = 0\) \cite{Carrozza:2015adg}, and extended by Klebanov and Tarnopolsky to the fermionic \(d = 1\) case \cite{Klebanov:2016xxf}. This model is similar to the Gurau-Witten model, with the difference being that there is 1/4 of the degrees of freedom. The model is likewise dominated by melon diagrams. The interaction term of the model is given below

\begin{equation}
V = \frac{1}{4}g\psi^{a_1b_1c_1}\psi^{a_1b_2c_2}\psi^{a_2b_1c_2}\psi^{a_2b_2c_1}
\end{equation}

Klebanov and Tarnopolsky also studied the same model, but in the case of complex fermions, for which the interaction term was the following

\begin{equation}
V = \frac{1}{4}g\Big(\bar\psi^{a_1b_1c_1}\bar\psi^{a_1b_2c_2}\bar\psi^{a_2b_1c_2}\bar\psi^{a_2b_2c_1}+\psi^{a_1b_1c_1}\psi^{a_1b_2c_2}\psi^{a_2b_1c_2}\psi^{a_2b_2c_1}\Big)
\end{equation}

This complex bipartite model with $O(N)^3$ symmetry, which is similar to the model in \cite{Gurau:2016lzk}, was first studied in \cite{Klebanov:2018fzb}. Complex tensor models of different symmetries have also been studied. The following model studied in \cite{Klebanov:2018fzb} has a $SU(N) \times O(N) \times SU(N) \times U(1)$ symmetry

\begin{equation}
V = \frac{g}{2}\psi^{a_1b_1c_1} \bar\psi^{a_1b_2c_2} \psi^{a_2b_1c_2} \bar\psi^{a_2b_2c_1}
\end{equation}

\newpage
\section{The Two Flavour Tensor Model}
The interaction term of the two flavour tensor model is defined as follows

\begin{equation}
\begin{split}
\MoveEqLeft
H = \frac{\beta g}{2}\{\psi_1^{a_1 b_1 c_1} \psi_1^{a_1 b_2 c_2} \psi_2^{a_2 b_1 c_2} \psi_2^{a_2 b_2 c_1} - \psi_1^{a_1 b_1 c_1} \psi_1^{a_2 b_1 c_2} \psi_2^{a_1 b_2 c_2} \psi_2^{a_2 b_2 c_1} + \psi_1^{a_1 b_1 c_1} \psi_1^{a_2 b_2 c_1} \psi_2^{a_1 b_2 c_2} \psi_2^{a_2 b_1 c_2}\} \\&
+ \frac{g}{4}\{\psi_1^{a_1 b_1 c_1} \psi_1^{a_1 b_2 c_2} \psi_1^{a_2 b_1 c_2} \psi_1^{a_2 b_2 c_1} + \psi_2^{a_1 b_1 c_1} \psi_2^{a_1 b_2 c_2} \psi_2^{a_2 b_1 c_2} \psi_2^{a_2 b_2 c_1}\}
\end{split}
\label{Hamiltonian}
\end{equation}

It is of note that this Hamiltonian is invariant under the transformation

\begin{equation}
\psi_1^{abc} \rightarrow A^{a}_{a'}B^{b}_{b'}C^{c}_{c'}\psi_i^{a'b'c'}
\end{equation}

Here, \(A\), \(B\), and \(C\) are orthogonal matrices, so the Hamiltonian has an \(O(N)^3\) symmetry.

A slightly wordier version of the Hamiltonian is as follows. First let us define

\begin{equation}
I_1(a_i, b_i, c_i) = \delta_{a_1 a_2}\delta_{b_1 b_3}\delta_{c_1 c_4}\delta_{b_2 b_4}\delta_{c_2 c_3}\delta_{a_3 a_4} - \delta_{a_1 a_3}\delta_{b_1 b_2}\delta_{c_1 c_4}\delta_{a_2 a_4}\delta_{c_2 c_3}\delta_{b_3 b_4} + \delta_{a_1 a_3}\delta_{b_1 b_4}\delta_{c_1 c_2}\delta_{a_2 a_4}\delta_{b_2 b_3}\delta_{c_3 c_4}
\end{equation}
\begin{equation}
I_2(a_i, b_i, c_i) = \delta_{a_1 a_2}\delta_{b_1 b_3}\delta_{c_1 c_4}\delta_{b_2 b_4}\delta_{c_2 c_3}\delta_{a_3 a_4}
\end{equation}

Then, the Hamiltonian can be written as

\begin{equation}
\begin{split}
\MoveEqLeft
H = \frac{\beta g}{2}I_1(a_i, b_i, c_i)\psi_1^{a_1b_1c_1}\psi_1^{a_2b_2c_2}\psi_2^{a_3b_3c_3}\psi_2^{a_4b_4c_4} \\&\qquad\qquad\qquad + \frac{g}{4}I_2(a_i, b_i, c_i)(\psi_1^{a_1b_1c_1}\psi_1^{a_2b_2c_2}\psi_1^{a_3b_3c_3}\psi_1^{a_4b_4c_4} + \psi_2^{a_1b_1c_1}\psi_2^{a_2b_2c_2}\psi_2^{a_3b_3c_3}\psi_2^{a_4b_4c_4})
\end{split}
\end{equation}

\subsection{The Propagator}
\begin{figure}[htb]
  \begin{center}  
    \includegraphics [width=1\textwidth, angle=0.]{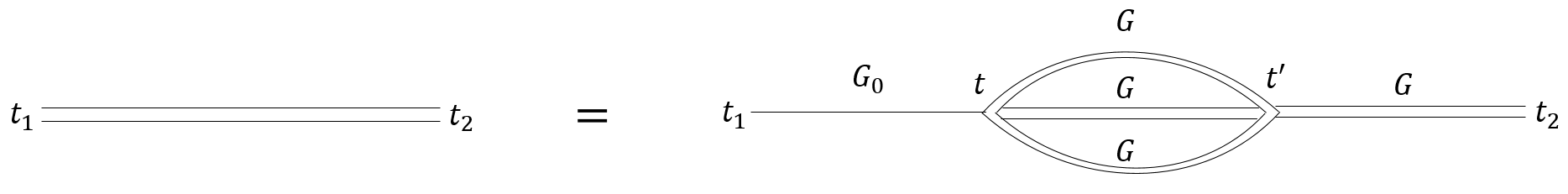}
  \end{center}
  \caption{Schwinger-Dyson equation. Double lines indicate the dressed propagators, and single lines indicate undressed propagators.}
  \label{fig:SD}
\end{figure}

The propagator of the two flavour tensor model can be calculated through the Schwinger-Dyson equation, similar to how we calculated the propagator for the SYK model. The Schwinger-Dyson equation is depicted graphically in figure \ref{fig:SD}.

The Schwinger-Dyson equation can be written down as

\begin{equation}
\begin{split}
G(t_2 - t_1) = G_0(t_2-t_1) + g^2(3\beta^2+1)N^3\int{dt dt' \ G_0(t-t_1)G(t'-t)^3 G(t_2 - t')}
\end{split}
\end{equation}

At large interactions, the first term can be ignored, and this gives us

\begin{equation}
\begin{split}
G(t_2 - t_1) = g^2(3\beta^2+1)N^3\int{dt dt' \ G_0(t-t_1)G(t'-t)^3 G(t_2 - t')}
\end{split}
\end{equation}

Since \(G_0(t_2-t_1) = \frac{\sgn(t_2-t_1)}{|t_2-t_1|^{1/2}}\), we find the propagator to be the following

\begin{equation}
\begin{split}
G(t_2 - t_1) = -\left(\frac{1}{4\pi(3\beta^2+1)g^2N^3}\right)^{\frac{1}{4}}\frac{\sgn(t_2-t_1)}{|t_2-t_1|^{1/2}}
\end{split}
\end{equation}

\newpage
\subsection{Four Point Functions}
Here, we repeat the process that we went through in the calculation of the four point functions of the SYK model in section \ref{SYK4}. We do not compute the full four point function, but rather exploit them to find the spectra.

\subsubsection{Four Point Functions and the Scaling Dimensions of \(\braket{\psi_1^{abc}\psi_1^{abc}\psi_1^{a'b'c'}\psi_1^{a'b'c'}}\) and \(\braket{\psi_1^{abc}\psi_1^{abc}\psi_2^{a'b'c'}\psi_2^{a'b'c'}}\)}
As shown previously, the four point functions can be found by computing the kernel of the ladder diagram, i.e., the individual ladders. There are two types of kernels for the four point functions \(\braket{\psi_1^{abc}(t_1)\psi_1^{abc}(t_2)\psi_1^{a'b'c'}(t_3)\psi_1^{a'b'c'}(t_4)}\), \(\braket{\psi_1^{abc}(t_1)\psi_1^{abc}(t_2)\psi_2^{a'b'c'}(t_3)\psi_2^{a'b'c'}(t_4)}\), and \(\braket{\psi_2^{abc}(t_1)\psi_2^{abc}(t_2)\psi_2^{a'b'c'}(t_3)\psi_2^{a'b'c'}(t_4)}\). Let us deal with the first two four point functions to simplify the argument. The last four point function is identical to the first, and hence without loss of generality, one can consider only the first two four point functions.

\begin{figure}[htb]
  \begin{center}  
    \includegraphics [width=1\textwidth, angle=0.]{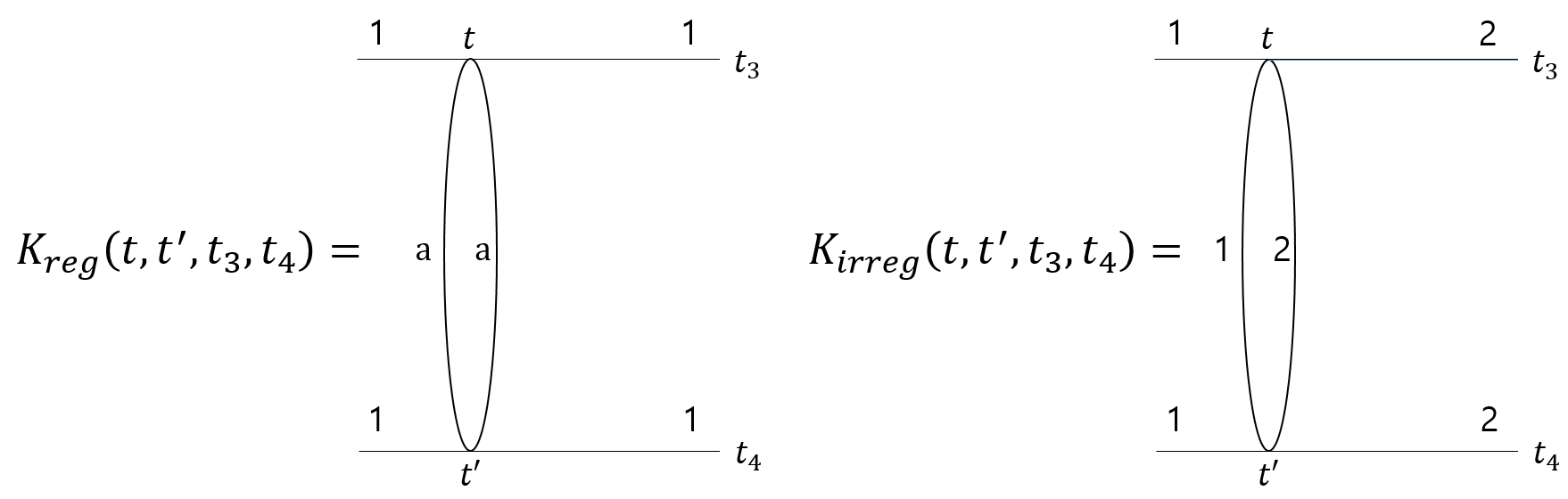}
  \end{center}
  \caption{The kernels of the four point function \(\braket{\psi_1^{abc}(t_1)\psi_1^{abc}(t_2)\psi_2^{a'b'c'}(t_3)\psi_2^{a'b'c'}(t_4)}\) and \(\braket{\psi_2^{abc}(t_1)\psi_2^{abc}(t_2)\psi_2^{a'b'c'}(t_3)\psi_2^{a'b'c'}(t_4)}\).}
  \label{fkernel}
\end{figure}

The two kernels are illustrated in figure \ref{fkernel}. \(K_{reg}\) does not change the flavour of what comes from the left. That is, if the particles that were coming from the left were \(\psi_1\)s, then after going through the \(K_{reg}\) kernel, the particles that come out would still be \(\psi_1\)s. On the other hand, \(K_{irreg}\) switches the flavour of the particles as they go through them. For example, if the particles coming from the left were \(\psi_1\)s, then the particles going out to the right would be \(\psi_2\)s.

Let us calculate the two different types of kernels. The kernel is formed from two insertions of the Hamiltonian

\begin{equation}
\begin{split}
\MoveEqLeft
K = -\Big\{\frac{\beta g^2}{4}I_1(a_i, b_i, c_i)I_1(a'_i, b'_i, c'_i)\psi_1^{a_1 b_1 c_1}(t) \psi_1^{a_2 b_2 c_2}(t) \psi_2^{a_3 b_3 c_3}(t) \psi_2^{a_4 b_4 c_4}(t) \psi_1^{a'_1 b'_1 c'_1}(t') \psi_1^{a'_2 b'_2 c'_2}(t') \\& \qquad \psi_2^{a'_3 b'_3 c'_3}(t') \psi_2^{a'_4 b'_4 c'_4}(t') + \frac{g^2}{16}I_2(a_i, b_i, c_i)I_2(a'_i, b'_i, c'_i)\psi_1^{a_1 b_1 c_1}(t) \psi_1^{a_2 b_2 c_2}(t) \psi_1^{a_3 b_3 c_3}(t) \psi_1^{a_4 b_4 c_4}(t) \\& \qquad \psi_1^{a'_1 b'_1 c'_1}(t') \psi_1^{a'_2 b'_2 c'_2}(t') \psi_1^{a'_3 b'_3 c'_3}(t') \psi_1^{a'_4 b'_4 c'_4}(t') + \frac{g^2}{16}I_2(a_i, b_i, c_i)I_2(a'_i, b'_i, c'_i)\psi_2^{a_1 b_1 c_1}(t) \psi_2^{a_2 b_2 c_2}(t) \\& \qquad \psi_2^{a_3 b_3 c_3}(t) \psi_2^{a_4 b_4 c_4}(t)\psi_2^{a'_1 b'_1 c'_1}(t') \psi_2^{a'_2 b'_2 c'_2}(t') \psi_2^{a'_3 b'_3 c'_3}(t') \psi_2^{a'_4 b'_4 c'_4}(t')\Big\} \psi_2^{a'b'c'}(t_3)\psi_2^{a'b'c'}(t_4)
\end{split}
\end{equation}

Note that we set the flavour of the external lines as 1 without loss of generality, and that the terms in which \(I_1\) and \(I_2\) are multiplied together are ignored, for their contractions lead to tadpole diagrams.

\(I_1 I_1\) interactions give the following term for \(K\), with Wick contractions. Note that we contract \(a_4b_4c_4\) and \(a'_4b'_4c'_4\) with \(a'b'c'\), which gives us \(\delta_{a_4a'_4}\delta_{b_4b'_4}\delta_{c_4c'_4}\). Other contractions of \(a'b'c'\) with \(a_ib_ic_i\) and \(a'_ib'_ic'_i\) give the same amplitude due to symmetry under the interchange of the position of \(\psi_2^{a_ib_ic_i}\), and hence can be accounted for by multiplying the resulting kernel from the contraction of \(a_4b_4c_4\) and \(a'_4b'_4c'_4\) with \(a'b'c'\) by 4

\begin{equation}
\begin{split}
\MoveEqLeft
\frac{\beta^2 g^2}{4}I_1(a_i, b_i, c_i)I_1(a'_i, b'_i, c'_i)\psi_1^{a_1 b_1 c_1}(t) \psi_1^{a_2 b_2 c_2}(t) \psi_2^{a_3 b_3 c_3}(t) \psi_2^{a_4 b_4 c_4}(t)\\& \qquad \psi_1^{a'_1 b'_1 c'_1}(t') \psi_1^{a'_2 b'_2 c'_2}(t') \psi_2^{a'_3 b'_3 c'_3}(t') \psi_2^{a'_4 b'_4 c'_4}(t') \psi_2^{a'b'c'}(t_3)\psi_2^{a'b'c'}(t_4) \\&
= \beta^2g^2\Big(\delta_{a_1 a_2}\delta_{b_1 b_3}\delta_{c_1 c_4}\delta_{b_2 b_4}\delta_{c_2 c_3}\delta_{a_3 a_4} - \delta_{a_1 a_3}\delta_{b_1 b_2}\delta_{c_1 c_4}\delta_{a_2 a_4}\delta_{c_2 c_3}\delta_{b_3 b_4} + \delta_{a_1 a_3}\delta_{b_1 b_4}\delta_{c_1 c_2}\delta_{a_2 a_4}\delta_{b_2 b_3}\delta_{c_3 c_4}\Big) \\& \qquad
\Big(\delta_{a'_1 a'_2}\delta_{b'_1 b'_3}\delta_{c'_1 c'_4}\delta_{b'_2 b'_4}\delta_{c'_2 c'_3}\delta_{a'_3 a'_4} - \delta_{a'_1 a'_3}\delta_{b'_1 b'_2}\delta_{c'_1 c'_4}\delta_{a'_2 a'_4}\delta_{c'_2 c'_3}\delta_{b'_3 b'_4} + \delta_{a'_1 a'_3}\delta_{b'_1 b'_4}\delta_{c'_1 c'_2}\delta_{a'_2 a'_4}\delta_{b'_2 b'_3}\delta_{c'_3 c'_4}\Big)\\& \qquad
\delta_{a_4 a'_4}\delta_{b_4 b'_4}\delta_{c_4 c'_4} \psi_1^{a_1b_1c_1}\psi_1^{a_2b_2c_2}\psi_2^{a_3b_3c_3}\psi_1^{a'_1b'_1c'_1}\psi_1^{a'_2b'_2c'_2}\psi_2^{a'_3b'_3c'_3}G(t_3-t)G(t_4-t') \\&
= \beta^2g^2\Big(\delta_{a_3a'_3}\delta_{b_2b'_2}\delta_{c_1c'_1}\delta_{a_1a_2}\delta_{b_1b_3}\delta_{c_2c_3}\delta_{a'_1a'_2}\delta_{b'_1b'_3}\delta_{c'_2c'_3} + \delta_{a_3a'_2}\delta_{b_2b'_1}\delta_{c_1c'_3}\delta_{a_1a_2}\delta_{b_1b_3}\delta_{c_2c_3}\delta_{b'_1b'_2}\delta_{c'_1c'_3}\delta_{a'_2a'_3}
\\& \qquad\qquad + \delta_{a_3a'_1}\delta_{b_2b'_3}\delta_{c_1c'_2}\delta_{a_1a_2}\delta_{b_1b_3}\delta_{c_2c_3}\delta_{c'_1c'_2}\delta_{a'_1a'_3}\delta_{b'_2b'_3} + \textrm{cyclic permutations of a, b, c}\Big)
\\&\qquad \psi_1^{a_1b_1c_1}\psi_1^{a_2b_2c_2}\psi_2^{a_3b_3c_3}\psi_1^{a'_1b'_1c'_1}\psi_1^{a'_2b'_2c'_2}\psi_2^{a'_3b'_3c'_3}G(t_3-t)G(t_4-t')
\end{split}
\end{equation}

Now let us perform the second set of wick contractions. Wick contraction between \(\psi_1^{a_1b_1c_1}\) and \(\psi_1^{a'_2b'_2c'_2}\) or \(\psi_1^{a'_1b'_1c'_1}\) and \(\psi_1^{a_2b_2c_2}\) results in an amplitude of \(g^2N\). Consequently, this is ignored in the large N limit. An intuitive reason why this is so can be seen in the Appendix. Continuing with the calculations

\begin{equation}
\begin{split}
\MoveEqLeft
\qquad = \beta^2g^2\Big(\delta_{a_3a'_3}\delta_{b_2b'_2}\delta_{c_1c'_1}\delta_{a_1a_2}\delta_{b_1b_3}\delta_{c_2c_3}\delta_{a'_1a'_2}\delta_{b'_1b'_3}\delta_{c'_2c'_3} + \delta_{a_3a'_2}\delta_{b_2b'_1}\delta_{c_1c'_3}\delta_{a_1a_2}\delta_{b_1b_3}\delta_{c_2c_3}\delta_{b'_1b'_2}\delta_{c'_1c'_3}\delta_{a'_2a'_3}
\\& \qquad\qquad + \delta_{a_3a'_1}\delta_{b_2b'_3}\delta_{c_1c'_2}\delta_{a_1a_2}\delta_{b_1b_3}\delta_{c_2c_3}\delta_{c'_1c'_2}\delta_{a'_1a'_3}\delta_{b'_2b'_3} + \textrm{cyclic permutations of a, b, c}\Big)
\\&\qquad \Big\{\delta_{a_2a'_2}\delta_{b_2b'_2}\delta_{c_2c'_2}\delta_{a_3a'_3}\delta_{b_3b'_3}\delta_{c_3c'_3}\psi_1^{a_1b_1c_1}\psi_1^{a'_1b'_1c'_1} + \delta_{a_1a'_1}\delta_{b_1b'_1}\delta_{c_1c'_1}\delta_{a_3a'_3}\delta_{b_3b'_3}\delta_{c_3c'_3}\psi_1^{a_2b_2c_2}\psi_1^{a'_2b'_2c'_2}
\\&\qquad + \delta_{a_1a'_1}\delta_{b_1b'_1}\delta_{c_1c'_1}\delta_{a_2a'_2}\delta_{b_2b'_2}\delta_{c_2c'_2}\psi_2^{a_3b_3c_3}\psi_2^{a'_3b'_3c'_3}\Big\}G(t_3-t)G(t_4-t')G(t-t')^2 \\&
= 3\beta^2g^2N^3\Big\{\delta_{a_1a'_1}\delta_{b_1b'_1}\delta_{c_1c'_1}\psi_1^{a_1b_1c_1}\psi_1^{a'_1b'_1c'_1} + \delta_{a_2a'_2}\delta_{b_2b'_2}\delta_{c_2c'_2}\psi_1^{a_2b_2c_2}\psi_1^{a'_2b'_2c'_2} \\&\qquad\qquad\qquad + \delta_{a_3a'_3}\delta_{b_3b'_3}\delta_{c_3c'_3}\psi_2^{a_3b_3c_3}\psi_2^{a'_3b'_3c'_3}\Big\}G(t_3-t)G(t_4-t')G(t-t')^2
\end{split}
\end{equation}

These repetitive and numerous calculations can be simplified tremendously if one uses the graphical notation of the Hamiltonian as outlined in the Appendix. Using the graphical method outlined in the Appendix, we find that the terms in which \(I_2\) are multiplied with one another gives

\begin{equation}
3g^2N^3\psi_2^{abc}(t)\psi_2^{abc}(t')G(t-t')^2G(t_3-t)G(t_4-t)
\end{equation} 

Summing the \(I_1I_1\) and the \(I_2I_2\) terms together, we find that the kernel \(K\) is

\begin{equation}
3g^2N^3\left\{(\beta^2+1)\psi_2^{abc}(t)\psi_2^{abc}(t')+2\beta^2\psi_1^{abc}(t)\psi_1^{abc}(t')\right\}G(t-t')^2G(t_3-t)G(t_4-t)
\end{equation} 

Therefore, since we contracted with $\psi_2^{a'b'c'}$ from the left, we determine that \(K_{reg}\) and \(K_{irreg}\) are the following from their definitions

\begin{equation}
K_{reg}(t_1,t_2,t_3,t_4) = -3g^2(\beta^2+1)N^3G(t-t')^2G(t_3-t)G(t_4-t)
\end{equation}
\begin{equation}
K_{irreg}(t_1,t_2,t_3,t_4)= -6\beta^2g^2N^3G(t-t')^2G(t_3-t)G(t_4-t)
\end{equation}

Now, let us define \(\Gamma\) as below

\begin{equation}
\Gamma^{n}_{reg}(t_1,t_2,t_3,t_4)= \braket{\psi_1(t_1)\psi_1(t_2)\psi_1(t_3)\psi_1(t_4)} \textrm{'s $n$ runged ladder}
\end{equation}
\begin{equation}
\Gamma^{n}_{irreg}(t_1,t_2,t_3,t_4)= \braket{\psi_1(t_1)\psi_1(t_2)\psi_2(t_3)\psi_2(t_4)} \textrm{'s $n$ runged ladder}
\end{equation}

\begin{figure}[htb]
  \begin{center}  
    \includegraphics [width=1\textwidth, angle=0.]{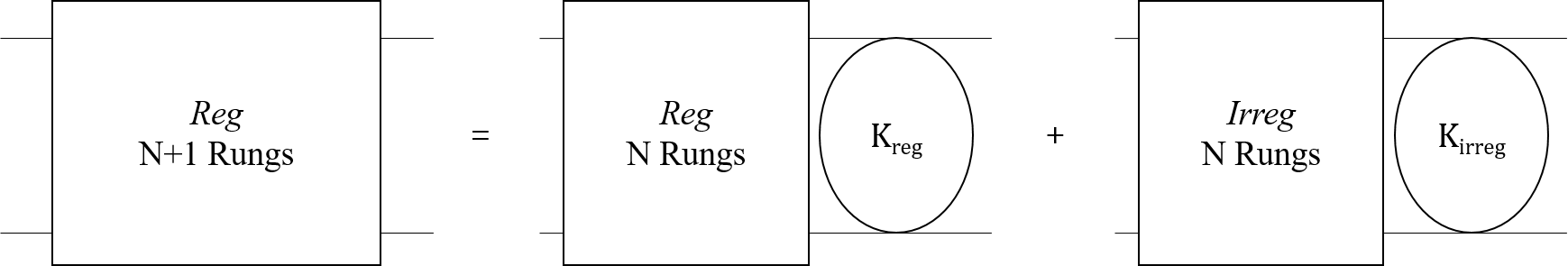}
    
    \vspace{0.3cm}

    \includegraphics [width=1\textwidth, angle=0.]{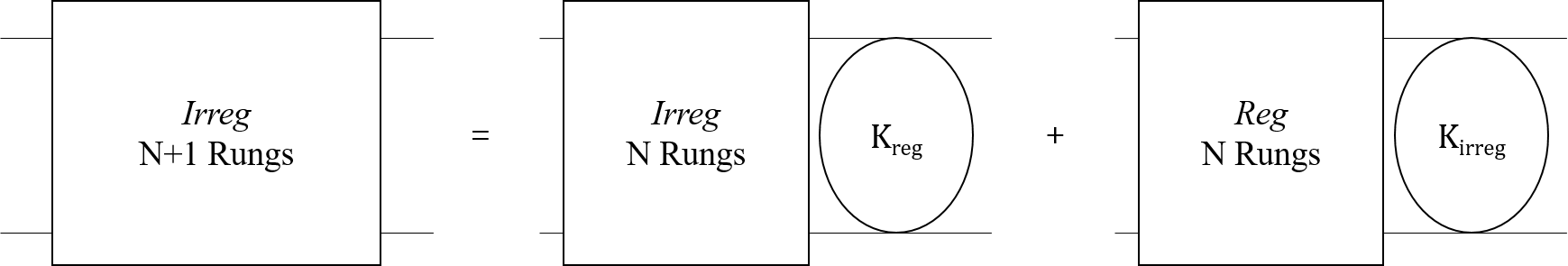}
  \end{center}
  \caption{The kernels of the four point function \(\braket{\psi_1^{abc}(t_1)\psi_1^{abc}(t_2)\psi_2^{a'b'c'}(t_3)\psi_2^{a'b'c'}(t_4)}\) and \(\braket{\psi_1^{abc}(t_1)\psi_1^{abc}(t_2)\psi_2^{a'b'c'}(t_3)\psi_2^{a'b'c'}(t_4)}\) satisfies the following equation with the ladder diagrams.}
  \label{kerneleqregirreg}
\end{figure}

Then, they would satisfy the following graphical equations as drawn in figure \ref{kerneleqregirreg}. This is equivalent to the following

\begin{equation}
\Gamma_{reg}^{N+1}(t_1,t_2,t_3,t_4) = \int{dt dt'}\Gamma_{reg}^{N}(t_1,t_2,t,t')K_{reg}(t,t',t_3,t_4)+\Gamma_{irreg}^{N}(t_1,t_2,t,t')K_{irreg}(t,t',t_3,t_4)
\end{equation}
\begin{equation}
\Gamma_{irreg}^{N+1}(t_1,t_2,t_3,t_4) = \int{dt dt'}\Gamma_{irreg}^{N}(t_1,t_2,t,t')K_{reg}(t,t',t_3,t_4)+\Gamma_{reg}^{N}(t_1,t_2,t,t')K_{irreg}(t,t',t_3,t_4)
\end{equation}

Consequently, the kernel can be written down as the following.

\begin{equation}
\begin{split}
\MoveEqLeft
-3g^2N^3
\begin{bmatrix}
\beta^2 +1 & 2\beta^2 \\ 2\beta^2 & \beta^2+1
\end{bmatrix}
\left(\frac{1}{4\pi(3\beta^2+1)g^2N^3}\right)\frac{\sgn(t_3-t)\sgn(t_4-t')}{|t_3-t|^{1/2}|t_4-t'|^{1/2}|t-t'|}
\end{split}
\end{equation}

Diagonalizing the kernel, the eigen-ladders come from the following four point functions

\begin{equation}
\braket{\psi_1^{abc}(t_1)\psi_1^{abc}(t_2)\psi_1^{a'b'c'}(t_3)\psi_1^{a'b'c'}(t_4) + \psi_1^{abc}(t_1)\psi_1^{abc}(t_2)\psi_2^{a'b'c'}(t_3)\psi_2^{a'b'c'}(t_4)}
\end{equation}
\begin{equation}
\braket{\psi_1^{abc}(t_1)\psi_1^{abc}(t_2)\psi_1^{a'b'c'}(t_3)\psi_1^{a'b'c'}(t_4) - \psi_1^{abc}(t_1)\psi_1^{abc}(t_2)\psi_2^{a'b'c'}(t_3)\psi_2^{a'b'c'}(t_4)}
\end{equation}.

Each has a corresponding eigen-kernel of

\begin{equation}
K_1 = -\frac{3}{4\pi}\frac{\sgn(t_3-t)\sgn(t_4-t')}{|t_3-t|^{1/2}|t_4-t'|^{1/2}|t-t'|}
\end{equation}
\begin{equation}
K_2 = -\frac{3}{4\pi}\frac{-\beta^2+1}{3\beta^2+1}\frac{\sgn(t_3-t)\sgn(t_4-t')}{|t_3-t|^{1/2}|t_4-t'|^{1/2}|t-t'|}
\end{equation}

Now, taking \(t_1\) and \(t_2\) to \(t_0\), let us define \(T_1\) and \(T_2\) to be the following

\begin{equation}
T_1(t_0, t_3, t_4) = \lim_{t_1, t_2 \rightarrow t_0}\braket{\psi_1^{abc}(t_1)\psi_1^{abc}(t_2)\psi_1^{a'b'c'}(t_3)\psi_1^{a'b'c'}(t_4) + \psi_1^{abc}(t_1)\psi_1^{abc}(t_2)\psi_2^{a'b'c'}(t_3)\psi_2^{a'b'c'}(t_4)}
\end{equation}
\begin{equation}
T_2(t_0, t_3, t_4) = \lim_{t_1, t_2 \rightarrow t_0}\braket{\psi_1^{abc}(t_1)\psi_1^{abc}(t_2)\psi_1^{a'b'c'}(t_3)\psi_1^{a'b'c'}(t_4) - \psi_1^{abc}(t_1)\psi_1^{abc}(t_2)\psi_2^{a'b'c'}(t_3)\psi_2^{a'b'c'}(t_4)}
\end{equation}

Then, the following equations hold

\begin{equation}
\begin{split}
\MoveEqLeft
T_1(t_0,t_1,t_2) = -G(t_0-t_1)G(t_0-t_2)+G(t_2-t_0)G(t_0-t_1) \\&\qquad\qquad\qquad + \int{dtdt'} T_1(t_0,t,t')K_1(t, t', t_1, t_2)
\end{split}
\end{equation}

\begin{equation}
\begin{split}
\MoveEqLeft
T_2(t_0,t_1,t_2) = G(t_0-t_1)G(t_0-t_2)-G(t_2-t_0)G(t_0-t_1) \\&\qquad\qquad\qquad + \int{dtdt'}T_2(t_0,t,t')K_2(t, t', t_1, t_2)
\end{split}
\end{equation}

Now, \(T_1\) and \(T_2\) are three point functions. Also, at large interactions, we can ignore the non-integral term. Therefore writing \(T_{1}(t_0, t_1, t_2)\) and \(T_{2}(t_0, t_1, t_2)\) as \(v_{1}(t_0, t_1, t_2)\), and \(v_{2}(t_0, t_1, t_2)\) for this limit we find

\begin{equation}
\begin{split}
\MoveEqLeft
\qquad g_{1}(h)v_{1}(t_0, t_1, t_2) = \int{dt dt'} K_{1}(t, t', t_1, t_2)v_{1}(t_0, t, t') \\&
g_{2}(h)v_{1}(t_0, t_1, t_2) = \int{dt dt'} K_{2}(t, t', t_1, t_2)v_{1}(t_0, t, t')
\end{split}
\end{equation}

Since the theory is conformal in this limit

\begin{equation}
v_{1,2}(t_0, t_1, t_2) = \frac{\sgn(t_1-t_2)}{|t_0-t_1|^h|t_0-t_2|^h|t_1-t_2|^{1/2-h}}
\end{equation}

where \(h\) is the scaling dimension of the following operators \(T_1\) and \(T_2\).

The \(SL(2)\) invariance lets us take \(t_0\) to infinity, and consequently we can just consider \(v_{1,2}(t_0, t_1, t_2) = \frac{\sgn(t_1-t_2)}{|t_1-t_2|^{1/2-h}}\). The eigenvalues to these eigenfunctions are

\begin{equation}
g_1(h) = -\frac{3}{2}\frac{\tan(\frac{\pi}{2}(h-1/2))}{h-1/2}
\end{equation}

\begin{equation}
g_2(h)= -\frac{3}{2}\frac{-\beta^2+1}{3\beta^2+1}\frac{\tan(\frac{\pi}{2}(h-1/2))}{h-1/2}
\end{equation}

Since the sum of ladder diagrams is \(\frac{1}{1-K}\), its amplitude is dominated by eigenvalues with \(g(h) = 1\). Furthermore, \(T_1\) and \(T_2\) are composed of the of the following conformal primaries

\begin{equation}
O^{2n+1}_{1} = \psi_1\partial_t^{2n+1}\psi_1 + \psi_2\partial_t^{2n+1}\psi_2
\end{equation}
\begin{equation}
O^{2n+1}_{2} = \psi_1\partial_t^{2n+1}\psi_1 - \psi_2\partial_t^{2n+1}\psi_2
\end{equation}

This is because with the Taylor expansion of \(T_1\) and \(T_2\), it is found that \(T\) is composed of \(\psi_1\partial_t^{n}\psi_1 \pm \psi_2\partial_t^{n}\psi_2\). However, when \(n = 2k\), the operators are not conformal primaries, and hence can be ignored.

\begin{figure}[htb]
  \begin{center}  
    \includegraphics [width=0.45\textwidth, angle=0.]{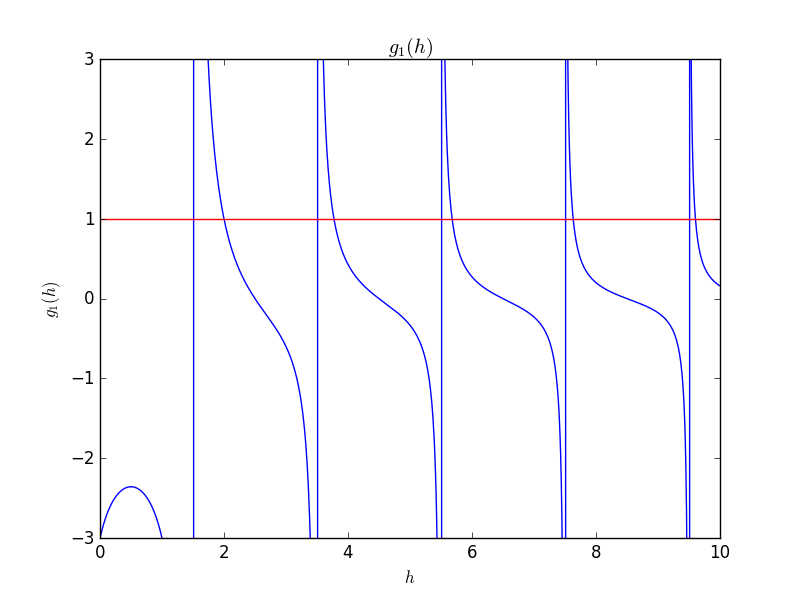}
    \includegraphics [width=0.45\textwidth, angle=0.]{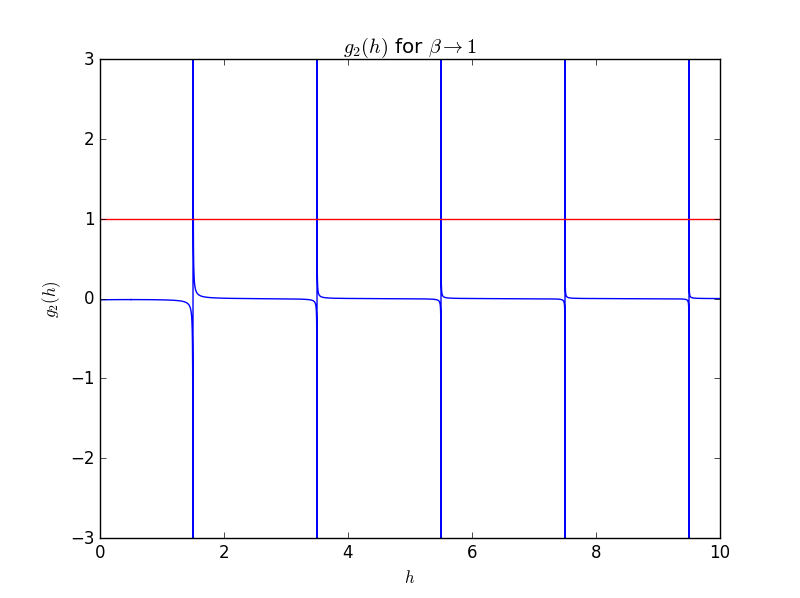}
  \end{center}
  \caption{Left: \(g_1(h)\) for any \(\beta\).    Right: \(g_2(h)\) for \(\beta \rightarrow 1\)}
  \label{g_1g_2}
\end{figure}

Consequently, the scaling dimensions of the bilinear conformal primary operators \(O_1^{2n+1}\) and \(O_2^{2n+1}\) are \(h\) that satisfy \(g_1(h) = 1\) and \(g_2(h) = 1\). Let us apply this method in the special case of \(\beta = 1\). \(g_1(h)\) and \(g_2(h)\) are as depicted in figure \ref{g_1g_2}.

The operators \(\psi_1\partial_t^{2n+1}\psi_1 + \psi_2\partial_t^{2n+1}\psi_2\) correspond to \(g_1\). Since \(g_1(h) = 1\) at \(h =\) 2.00, 3.77, 5.68, 7.63, 9.60 \(\dots\), the scaling dimensions are taken to be 2.00, 3.77, 5.68, 7.63, 9.60 and so on. It eventually converges to \(2n+\frac{3}{2}\).

Similarly, the operators \(\psi_1\partial_t^{2n+1}\psi_1 - \psi_2\partial_t^{2n+1}\psi_2\) correspond to \(g_2\). Here, the \(h\) in which \(g_2(h) = 1\) is dependent on \(\beta\). The situation in which \(\beta = \pm 1\) is particularly interesting: since the coefficient is zero, it seems that there are no solutions. This is not completely true due to the poles from the tangent function. This behavior can be studied by taking the limit of \(\beta \rightarrow \pm 1\).

For \(\beta \rightarrow 1\), we find that the solutions are \(h = \) 1.5, 3.5, 5.5, 7.5, 9.5 \(\dots\). Hence, the scaling dimension of the operators is \(2n+\frac{3}{2}\). This perfect behavior is due fact that in the \(\beta \rightarrow 1\) limit the tangent function in \(g_2\) must diverge in order to cancel out the coefficient which goes to zero. An intuitive reason behind this phenomenon in the limit of \(\beta \rightarrow 1\) is further discussed in section \ref{alpha1}.

Note that \(g_1(h)\) is analogous to the flavourless real fermion model \cite{Klebanov:2016xxf}. A sanity check is taking the \(\beta = 0\). In this limit the \(g_2(h)\) spectrum converges to the \(g_1(h)\) spectrum, and this is as expected: since the two flavours are decoupled at \(\beta = 0\), we expect the theory to be identical to \cite{Klebanov:2016xxf}.

Further proof of this result comes from Gross and Rosenhaus's results for the SYK model at $f = 2$ \cite{Gross:2016kjj}. Their result is that

\begin{align}
&K^{kk}(\tau_i) = - J^2\frac{q_k - 1}{K_k Q_k}G_k(\tau_{13})G_k(\tau_{24})\frac{1}{G_k(\tau_{34})^2}\prod_{a=1}^{f}G_a(\tau_{34})^{q_a} \\
&K^{kl}(\tau_i) = - J^2\frac{q_l}{K_k Q_k}G_k(\tau_{13})G_k(\tau_{24})\frac{1}{G_k(\tau_{34})G_l(\tau_{34})}\prod_{a=1}^{f}G_a(\tau_{34})^{q_a}
\end{align}

Where \(q_a\) is the number of times the the \(a\)th fermion appears in the interaction term. \(Q_k = \prod_{k \neq a}{q_k}\), and \(K_k = \frac{N_k}{\sum{N_i}}\). Letting \(q_1 = 2\), \(q_2 = 2\), \(N_1 = N\), and \(N_2 = N\)

\begin{align}
&K^{kk} = -J^2G_k(\tau_{13})G_k(\tau_{24})G_l(\tau_{34})^2 \\
&K^{kl} = - 2J^2G_k(\tau_{13})G_k(\tau_{24})G_k(\tau_{34})G_l(\tau_{34})\\
&K(\tau_i) = -J^2 \begin{bmatrix}
1 & 2 \\ 2 & 1
\end{bmatrix}
\left(\frac{b_k}{J^{2\Delta_k}}\right)^4G(\tau_{13})G(\tau_{24})G(\tau_{34})^2    
\end{align}

Where \(\prod_{a=1}^f b_a^{q_a} = \frac{1}{2\pi}(1-2\Delta_k) \tan(\pi\Delta_k)\). Plugging in \(\Delta = 1/4\)

\begin{equation}
\begin{split}
\MoveEqLeft
-\begin{bmatrix}
1 & 2 \\ 2 & 1
\end{bmatrix}
\left(\frac{1}{4\pi}\right)\frac{\sgn(\tau_1-\tau_3)\sgn(\tau_2-\tau_4)}{|\tau_1-\tau_3|^{1/2}|\tau_2-\tau_4|^{1/2}|\tau_3-\tau_4|}
\end{split}
\end{equation}

This is identical to the kernel when \(\beta = \infty\) in our model.

\subsubsection{Four Point Functions and the Scaling Dimensions of \(\braket{\psi_1^{abc}\psi_2^{abc}\psi_1^{a'b'c'}\psi_2^{a'b'c'}}\) and \(\braket{\psi_1^{abc}\psi_2^{abc}\psi_2^{a'b'c'}\psi_1^{a'b'c'}}\)}
\begin{figure}[htb]
  \begin{center}  
    \includegraphics [width=1\textwidth, angle=0.]{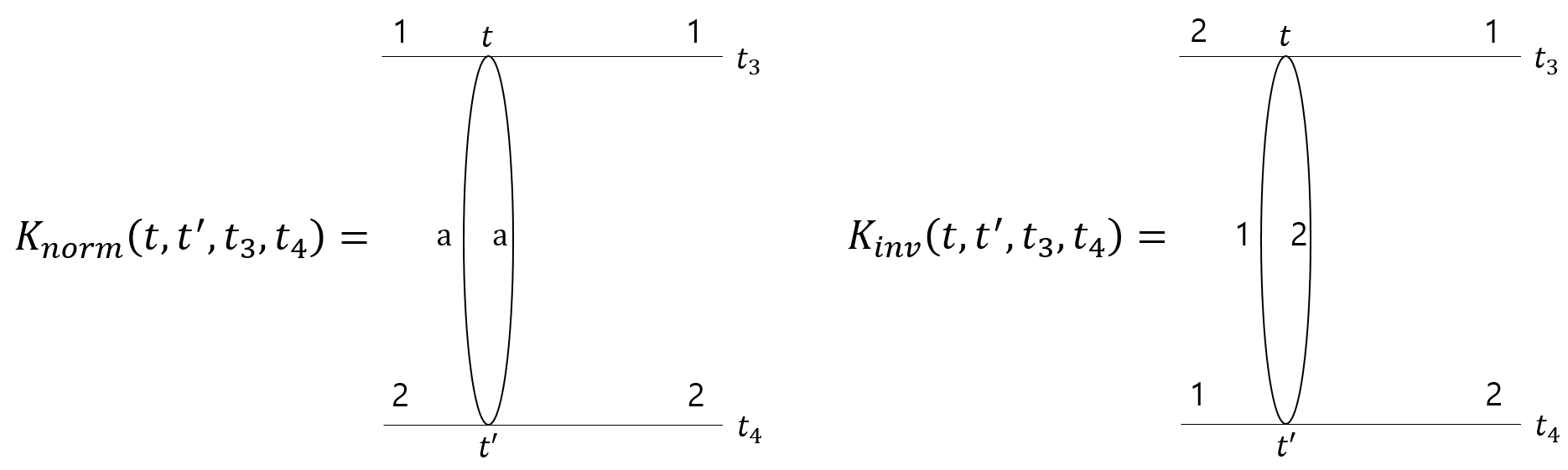}

  \end{center}
  \caption{The kernels of the four point function \(\braket{\psi_1^{abc}(t_1)\psi_2^{abc}(t_2)\psi_1^{a'b'c'}(t_3)\psi_2^{a'b'c'}(t_4)}\) and \(\braket{\psi_1^{abc}(t_1)\psi_2^{abc}(t_2)\psi_2^{a'b'c'}(t_3)\psi_1^{a'b'c'}(t_4)}\)}
  \label{fig:FPD}
\end{figure}

\(\braket{\psi_1^{abc}(t_1)\psi_2^{abc}(t_2)\psi_1^{a'b'c'}(t_3)\psi_2^{a'b'c'}(t_4)}\) and \(\braket{\psi_1^{abc}(t_1)\psi_2^{abc}(t_2)\psi_2^{a'b'c'}(t_3)\psi_1^{a'b'c'}(t_4)}\) can be computed in the same manner as in the previous section. The kernel for the two four point functions are as drawn in figure \ref{fig:FPD}. \(K_{norm}\) keeps the flavour order. For example, if \(\psi_1\) and \(\psi_2\) came in from the left, the \(\psi_1\) and \(\psi_2\) emerges from the right. On the other hand, \(K_{inv}\) switches the flavour order of what is coming from the left. That is, if \(\psi_1\) and \(\psi_2\) came in from the left, the \(\psi_2\) and \(\psi_1\) emerges from the right.

\(K_{inv}\) is the following

\begin{equation}
\begin{split}
\MoveEqLeft
-\frac{\beta^2g^2}{4}I_1(a_i, b_i, c_i)I_1(a'_i, b'_i, c'_i)\psi_1^{a_1 b_1 c_1}(t) \psi_1^{a_2 b_2 c_2}(t) \psi_2^{a_3 b_3 c_3}(t) \psi_2^{a_4 b_4 c_4}(t)\\& \psi_1^{a'_1 b'_1 c'_1}(t') \psi_1^{a'_2 b'_2 c'_2}(t') \psi_2^{a'_3 b'_3 c'_3}(t') \psi_2^{a'_4 b'_4 c'_4}(t')\psi_1^{a'b'c'}(t_3)\psi_2^{a'b'c'}(t_4) \\&
= 6\beta^2g^2N^3\psi_1^{abc}(t)\psi_2^{abc}(t')G(t_3-t)G(t_4-t')G(t-t')^2
\end{split}
\end{equation}

\(K_{norm}\), on the other hand, is of the following form. The factor of 2 comes from the fact that although we only do the contraction of \(\psi_1\psi_1\psi_2\psi_2\psi_2\psi_2\psi_2\psi_2\), the contraction of \(\psi_1\psi_1\psi_1\psi_1\psi_1\psi_1\psi_2\psi_2\) should also be considered

\begin{equation}
\begin{split}
\MoveEqLeft
-\frac{\beta g^2}{8}\times2I_1(a_i, b_i, c_i)I_2(a'_i, b'_i, c'_i)\psi_1^{a_1 b_1 c_1}(t) \psi_1^{a_2 b_2 c_2}(t) \psi_2^{a_3 b_3 c_3}(t) \psi_2^{a_4 b_4 c_4}(t)\\& \psi_2^{a'_1 b'_1 c'_1}(t') \psi_2^{a'_2 b'_2 c'_2}(t') \psi_2^{a'_3 b'_3 c'_3}(t') \psi_2^{a'_4 b'_4 c'_4}(t')\psi_1^{a'b'c'}(t_3)\psi_2^{a'b'c'}(t_4) \\&
= 6 \beta g^2N^3 \psi_1^{a''b''c''}(t)\psi_2^{a''b''c''}(t') G(t_3-t)G(t_4-t') + o(N^2)
\end{split}
\end{equation}

Now, let us define $\Gamma$ as the following

\begin{equation}
\Gamma_{norm}^n = \braket{\psi_1(t_1)\psi_2(t_2)\psi_1(t_3)\psi_2(t_4)} \textrm{'s n runged ladder}
\end{equation}
\begin{equation}
\Gamma_{inv}^n = \braket{\psi_1(t_1)\psi_2(t_2)\psi_2(t_3)\psi_1(t_4)} \textrm{'s n runged ladder}
\end{equation}

\begin{figure}[htb]
  \begin{center}  
    \includegraphics [width=1\textwidth, angle=0.]{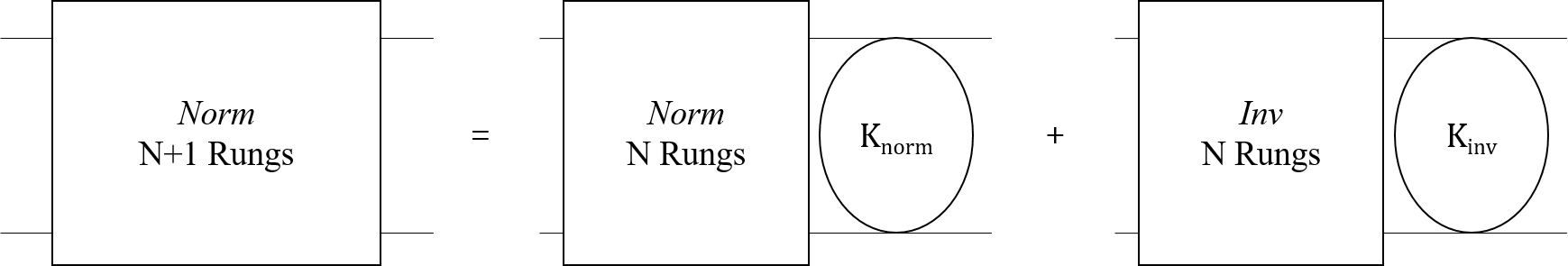}
    
    \vspace{0.3cm}
    
    \includegraphics [width=1\textwidth, angle=0.]{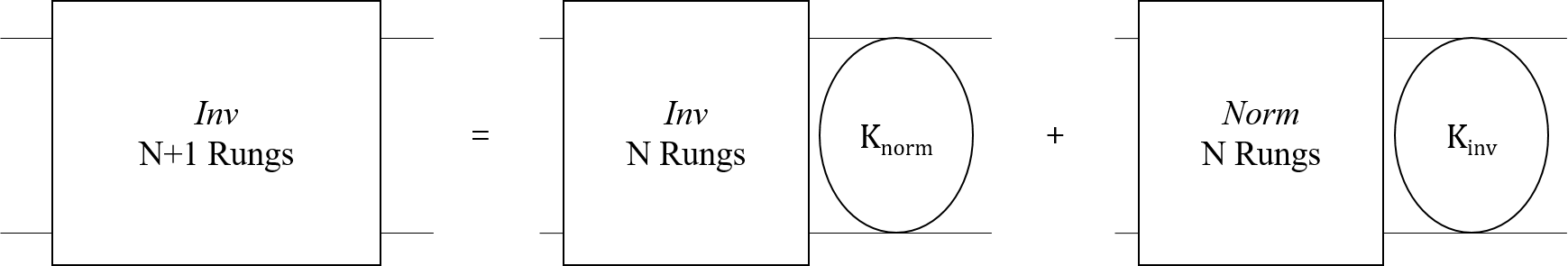}
  \end{center}
  \caption{The ladder diagram of \(\braket{\psi_1(t_1)\psi_2(t_2)\psi_1(t_3)\psi_2(t_4) \pm \psi_1(t_1)\psi_2(t_2)\psi_2(t_3)\psi_1(t_4)}\) with \(N+1\) rungs satisfy these relations with the \(N\) runged ladder diagrams.}
  \label{LERS}
\end{figure}

$\Gamma$ and $K$ satisfy the following graphical equations as drawn in figure \ref{LERS}. This is also equivalent to the following.

\begin{equation}
\Gamma^{N+1}_{norm}(t_1,t_2,t_3,t_4) = \int{dt dt'}\Gamma^{N}_{norm}(t_1,t_2,t,t')K_{norm}(t,t',t_3,t_4)+\Gamma^{N}_{inv}(t_1,t_2,t,t')K_{inv}(t,t',t_3,t_4)
\end{equation}
\begin{equation}
\Gamma^{N+1}_{inv}(t_1,t_2,t_3,t_4) = \int{dt dt'}\Gamma^{N}_{norm}(t_1,t_2,t,t')K_{inv}(t,t',t_3,t_4)+\Gamma^{N}_{inv}(t_1,t_2,t,t')K_{norm}(t,t',t_3,t_4)
\end{equation}

The kernel is therefore the following,

\begin{equation}
-6g^2N^3
\begin{bmatrix}
\beta & \beta^2\\
\beta^2 & \beta
\end{bmatrix}
\left(\frac{1}{4\pi(3\beta^2+1)g^2N^3}\right)\frac{\sgn(t_3-t)\sgn(t_4-t')}{|t_3-t|^{1/2}|t_4-t'|^{1/2}|t-t'|}
\end{equation}

Diagonalizing the kernel, the eigen-ladders come from the following four point functions

\begin{equation}
\braket{\psi_1(t_1)\psi_2(t_2)\psi_1(t_3)\psi_2(t_4) \pm \psi_1(t_1)\psi_2(t_2)\psi_2(t_3)\psi_1(t_4)}
\end{equation}

Each has a corresponding eigen-kernel of

\begin{equation}
K_3 = -\frac{3}{2\pi}\frac{\beta^2+\beta}{3\beta^2+1}\frac{\sgn(t_3-t)\sgn(t_4-t')}{|t_3-t|^{1/2}|t_4-t'|^{1/2}|t-t'|}
\end{equation}
\begin{equation}
K_4 = -\frac{3}{2\pi}\frac{-\beta^2+\beta}{3\beta^2+1}\frac{\sgn(t_3-t)\sgn(t_4-t')}{|t_3-t|^{1/2}|t_4-t'|^{1/2}|t-t'|}
\end{equation}

Just as we did in the previous section, we again take the limit of \(t_1, t_2 \rightarrow t_0\). Let us define \(T_3\) and \(T_4\) as the following

\begin{equation}
T_3(t_0, t_3, t_4) = \lim_{t_1, t_2 \rightarrow t_0}\braket{\psi_1(t_1)\psi_2(t_2)\psi_1(t_3)\psi_2(t_4) + \psi_1(t_1)\psi_2(t_2)\psi_2(t_3)\psi_1(t_4)}
\end{equation}
\begin{equation}
T_4(t_0, t_3, t_4) = \lim_{t_1, t_2 \rightarrow t_0}\braket{\psi_1(t_1)\psi_2(t_2)\psi_1(t_3)\psi_2(t_4) - \psi_1(t_1)\psi_2(t_2)\psi_2(t_3)\psi_1(t_4)}
\end{equation}

$T_3$ and $T_4$ satisfy the following equations

\begin{equation}
\begin{split}
\MoveEqLeft
T_3(t_0,t_1,t_2) = -\delta(t_0-t_1)\delta(t_0-t_2)+\delta(t_2-t_0)\delta(t_0-t_1)\\& \qquad\qquad\qquad+T_3(t_0,t,t')\Big\{K_{norm}(t, t', t_1, t_2)+K_{inv}(t, t', t_1, t_2)\Big\}
\end{split}
\end{equation}

\begin{equation}
\begin{split}
\MoveEqLeft
T_4(t_0,t_1,t_2) = -\delta(t_0-t_1)\delta(t_0-t_2)+\delta(t_2-t_0)\delta(t_0-t_1)\\& \qquad\qquad\qquad+T_4(t_0,t,t')\Big\{K_{norm}(t, t', t_1, t_2)-K_{inv}(t, t', t_1, t_2)\Big\}
\end{split}
\end{equation}

The first terms can be ignored again at large interactions. Thus writing down \(T_{3,4}(t_0, t_1, t_2)\) as \(v_{3,4}(t_0, t_1, t_2)\), we find

\begin{equation}
g_{3,4}(h)v_{3,4}(t_0,t_1,t_2) = \int{dtdt'}v_{3,4}(t_0,t,t')\Big\{K_{norm}(t, t', t_1, t_2) \pm K_{inv}(t, t', t_1, t_2)\Big\}
\end{equation}

For \(T_3(t_0, t_1, t_2)\), since it is symmetric under the interchange of \(t_1\) and \(t_2\), we can use \(v_3(t_0, t_1,t_2) = \frac{1}{|t_0-t_1|^h|t_0-t_2|^h|t_1-t_2|^{1/2-h}}\) as the eigenfunction. On the contrary, for \(T_4(t_0, t_1, t_2)\), we can use \(v_4(t_0, t_1, t_2) = \frac{\sgn(t_1-t_2)}{|t_0-t_1|^h|t_0-t_2|^h|t_1-t_2|^{1/2-h}}\), since it is anti-symmetric under the interchange of \(t_1\) and \(t_2\).

In addition, the \(SL(2)\) invariance lets us take \(t_0\) to infinity, and consequently we can just consider \(v_3(t_0, t_1, t_2) = \frac{1}{|t_1-t_2|^{1/2-h}}\) and \(v_4(t_0, t_1, t_2) = \frac{\sgn(t_1-t_2)}{|t_1-t_2|^{1/2-h}}\). The eigenvalues to these eigenfunctions are

\begin{equation}
g_3(h) = \frac{3\beta^2-3\beta}{(3\beta^2+1)}\frac{\tan(\pi h/2 + \pi/4)}{h-\frac{1}{2}}
\end{equation}

\begin{equation}
g_4(h) = -\frac{3\beta^2+3\beta}{(3\beta^2+1)}\frac{\tan(\pi h/2 - \pi/4)}{h-\frac{1}{2}}
\end{equation}

Since \(T_3\) and \(T_4\) correspond to \(O_3^{2n} = \psi_1 \partial_t^{2n} \psi_2 - \psi_2 \partial_t^{2n} \psi_1\) and \(O_4^{2n+1} = \psi_1 \partial_t^{2n+1} \psi_2 + \psi_2 \partial_t^{2n+1} \psi_1\), one can find the scaling dimensions of the above operators through finding when \(g_{3,4}(h) = 1\).

\begin{figure}[htb]
  \begin{center}  
    \includegraphics [width=0.45\textwidth, angle=0.]{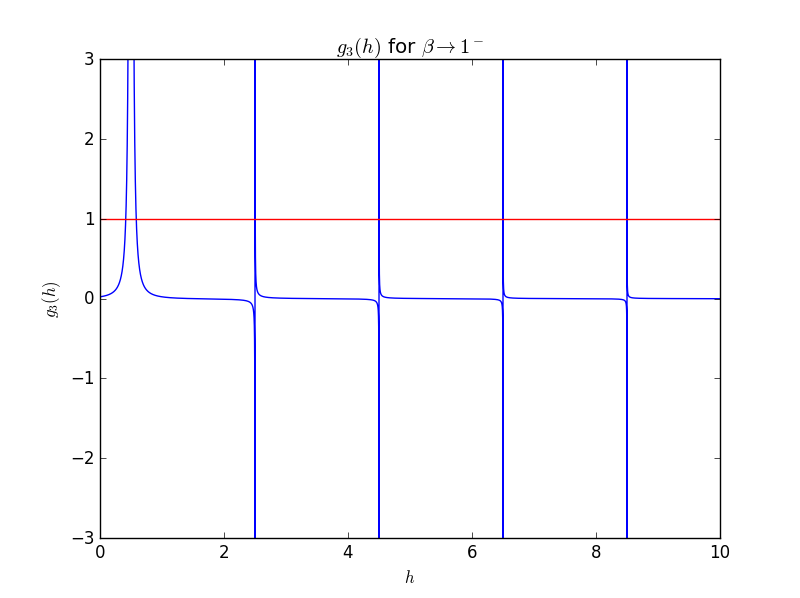}
    \includegraphics [width=0.45\textwidth, angle=0.]{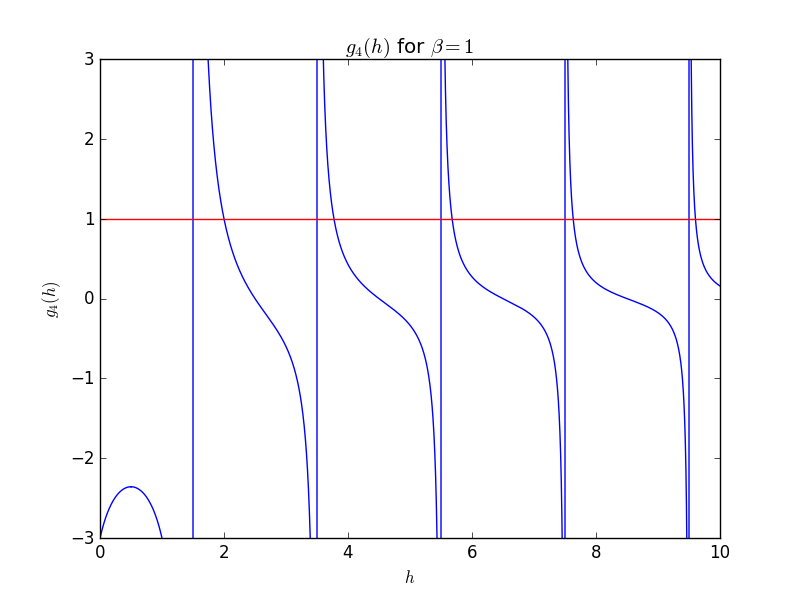}
  \end{center}
  \caption{Left: \(g_3(h)\) at \(\beta \rightarrow 1^-\), Right: \(g_4(h)\) at \(\beta = 1\).}
  \label{g_3g_4}
\end{figure}

Let us find the scaling dimensions of the operators at \(\beta = 1\). Note that since \(g_3 = 0\) at this value of \(\beta\), we find the scaling dimensions of the operators \(\psi_1 \partial_t^{2n} \psi_2 - \psi_2 \partial_t^{2n} \psi_1\) by taking the limit of \(\beta \rightarrow 1^-\). \(g_3(h)\) and \(g_4(h)\) are as illustrated in figure \ref{g_3g_4}.

\begin{figure}[htb]
  \begin{center}  
    \includegraphics [width=0.45\textwidth, angle=0.]{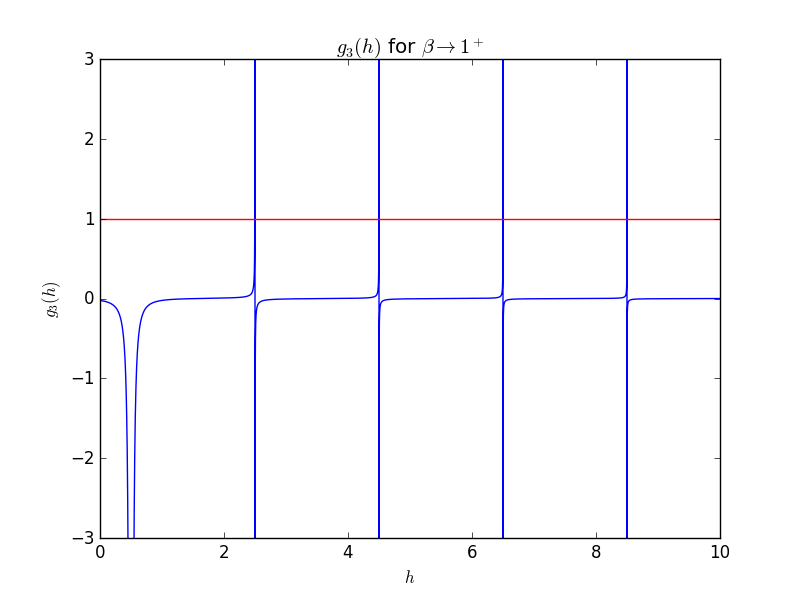}
  \end{center}
  \caption{\(g_3(h)\) at \(\beta \rightarrow 1^+\).}
  \label{g_3'}
\end{figure}

\(g_3(h)\) goes to 1 at 0.5, 2.5, 4.5, 6.5, 8.5, \(\dots\). Therefore, the scaling dimensions of \(\psi_1 \partial_t^{2n} \psi_2 - \psi_2 \partial_t^{2n} \psi_1\) are \(2n+\frac{1}{2}\). This happens since the coefficient \(\frac{3\beta^2-3\beta}{3\beta^2+1}\) goes to 0 as \(\beta\) approaches 1. For \(g_3(h)\) to be 1, \(\tan(\pi h/2 + \pi/4)\) needs to diverge to counteract the coefficient in front of it, and this only happens at \(h = 2n + 1/2\). Note that the pole around \(h = 0.5\) switches sign if \(\beta\) approaches 1 from above as in figure \ref{g_3'}. Thus, the scaling dimension near 0.5 does not exist anymore. The reason behind this disappearance is discussed later in section 2.4.

Similarly, \(g_4(h)\) goes to 1 at 2.00, 3.77, 5.68, 7.63, 9.60, \(\dots\), the scaling dimensions of \(\psi_1 \partial_t^{2n+1} \psi_2 + \psi_2 \partial_t^{2n+1} \psi_1\) are 2.00, 3.77, 5.68, 7.63, 9.60, \(\dots\), and it approaches \(2n+\frac{3}{2}\) with the increase of \(n\).

The reason why at \(\beta = 1\) the scaling dimensions of \(\psi_1 \partial_t^{2n} \psi_2 - \psi_2 \partial_t^{2n} \psi_1\) is exactly \(h = 2n + 1/2\) is interesting, and will be explained in the next section.

We can check this result by comparing it to that of some other models. The complex bipartite fermion model \cite{Gurau:2016lzk, Klebanov:2018fzb} has the Hamiltonian

\begin{equation}
\begin{split}
\MoveEqLeft
H = \frac{1}{4}g\Big(\psi^{a_1b_1c_1}\psi^{a_1b_2c_2}\psi^{a_2b_1c_2}\psi^{a_2b_2c_1}+\bar\psi^{a_1b_1c_1}\bar\psi^{a_1b_2c_2}\bar\psi^{a_2b_1c_2}\bar\psi^{a_2b_2c_1}\Big)
\end{split}
\end{equation}

Klebanov and Tarnopolsky found that on the symmetric sector
\begin{equation}
g_{sym}(h) = \frac{3}{2}\frac{\tan(\frac{\pi}{2}(h+\frac{1}{2}))}{h-1/2}
\end{equation}

Writing \(\psi^{abc}=\psi_1^{abc}+i\psi_2^{abc}\) and \(\bar\psi^{abc}=\psi_1^{abc}-i\psi_2^{abc}\) we find

\begin{equation}
\begin{split}
\MoveEqLeft
H = g\Big(\psi_1^{a_1b_1c_1}\psi_1^{a_1b_2c_2}\psi_2^{a_2b_1c_2}\psi_2^{a_2b_2c_1}+\psi_1^{a_1b_1c_1}\psi_2^{a_1b_2c_2}\psi_1^{a_2b_1c_2}\psi_2^{a_2b_2c_1}+\psi_1^{a_1b_1c_1}\psi_2^{a_1b_2c_2}\psi_2^{a_2b_1c_2}\psi_1^{a_2b_2c_1}\Big)
\\&\qquad-\frac{g}{2}\Big(\psi_1^{a_1b_1c_1}\psi_1^{a_1b_2c_2}\psi_1^{a_2b_1c_2}\psi_1^{a_2b_2c_1} + \psi_2^{a_1b_1c_1}\psi_2^{a_1b_2c_2}\psi_2^{a_2b_1c_2}\psi_2^{a_2b_2c_1}\Big)
\end{split}
\end{equation}

Therefore, the complex bipartite tensor model corresponds to \(\beta = -1\) in our Hamiltonian. Putting \(\beta = -1\) in the formula that we found for \(g_3(h)\)

\begin{equation}
g_3(h) = \frac{3}{2}\frac{\tan(\pi h/2 + \pi/4)}{h-\frac{1}{2}}
\end{equation}

\(g_3(h) = g_{sym}(h)\). Therefore, we see a match.

\subsubsection{$\beta = 1$ and the Scaling Dimensions of $\psi_1 \partial_t^{n} \psi_2$}
\label{alpha1}
At \(\beta = 1\), doing a transformation of \(\tilde\psi_1 = \frac{1}{\sqrt{2}}(\psi_1 + \psi_2)\) and \(\tilde\psi_2 = \frac{1}{\sqrt{2}}(\psi_1 - \psi_2)\) transforms the Hamiltonian as follows

\begin{equation}
H = \frac{g}{2}(\tilde\psi_1^{a_1b_1c_1}\tilde\psi_1^{a_1b_2c_2}\tilde\psi_1^{a_2b_1c_2}\tilde\psi_1^{a_2b_2c_1} + \tilde\psi_2^{a_1b_1c_1}\tilde\psi_2^{a_1b_2c_2}\tilde\psi_2^{a_2b_1c_2}\tilde\psi_2^{a_2b_2c_1})
\end{equation}

Similarly
\begin{equation}
\begin{split}
\MoveEqLeft
O_2^{2n+1} = \psi_1 \partial_t^{2n+1} \psi_1 - \psi_2 \partial_t^{2n+1} \psi_2 \\&\qquad = \tilde\psi_2 \partial_t^{2n+1} \tilde\psi_1 + \tilde\psi_1 \partial_t^{2n+1} \tilde\psi_2
\end{split}
\end{equation}
\begin{equation}
\begin{split}
\MoveEqLeft
O_3^{2n} = \psi_1 \partial_t^{2n} \psi_2 - \psi_2 \partial_t^{2n} \psi_1 \\&\qquad = \tilde\psi_2 \partial_t^{2n} \tilde\psi_1 - \tilde\psi_1 \partial_t^{2n} \tilde\psi_2
\end{split}
\end{equation}

Since \(\tilde\psi_1\) and \(\tilde\psi_2\) are decoupled, no anomalous dimensions occur in their products; also, the dimensions of the operators \(\tilde\psi_i\) are \(1/4\). Hence, \(\tilde\psi_2 \partial_t^{2n+1} \tilde\psi_1 + \tilde\psi_1 \partial_t^{2n+1} \tilde\psi_2\) simply takes the form \(\left(\frac{1}{\Lambda}\right)^{2n+3/2}\). Similarly, \(\tilde\psi_2 \partial_t^{2n} \tilde\psi_1 - \tilde\psi_1 \partial_t^{2n} \tilde\psi_2\) takes the form \(\left(\frac{1}{\Lambda}\right)^{2n+1/2}\). Hence, the scaling dimensions of \(O_2^{2n+1}\) is  \(2n + \frac{3}{2}\), and \(O_3^{2n}\), \(2n + \frac{1}{2}\).

In general, the scaling dimension of the operator \(\tilde\psi_1 \partial_t^{n} \tilde \psi_2\) is going to be \(n + \frac{1}{2}\), due to the decoupling of $\tilde \psi$ with one another.

\subsubsection{$\beta = -\frac{1}{3}$ and the Scaling Dimensions of $\tilde\psi_i \partial_t^{2n+1} \tilde\psi_i$}
Another special case is \(\beta = -\frac{1}{3}\). A transformation of \(\tilde\psi_1 = \frac{1}{\sqrt{2}}(\psi_1 + \psi_2)\) and \(\tilde\psi_2 = \frac{1}{\sqrt{2}}(\psi_1 - \psi_2)\) transforms the Hamiltonian as follows

\begin{equation}
H = \frac{g}{3}(\tilde\psi_1^{a_1b_1c_1}\tilde\psi_1^{a_1b_2c_2}\tilde\psi_2^{a_2b_1c_2}\tilde\psi_2^{a_2b_2c_1} + \tilde\psi_1^{a_1b_1c_1}\tilde\psi_2^{a_1b_2c_2}\tilde\psi_1^{a_2b_1c_2}\tilde\psi_2^{a_2b_2c_1} + \tilde\psi_1^{a_1b_1c_1}\tilde\psi_2^{a_1b_2c_2}\tilde\psi_2^{a_2b_1c_2}\tilde\psi_1^{a_2b_2c_1})
\end{equation}

This corresponds to \(\beta \rightarrow \infty\), and consequently would match the SYK model from \cite{Gross:2016kjj} at \(q_1 = 2, q_2 = 2, N_1 = N, N_2 = N\).

Gross and Rosenhaus found the spectra of \(\psi_i \partial_t^{2n+1} \psi_i\) to be the following

\begin{equation}
g_k(h)=\rho(h)\sigma_k
\end{equation}

where \(\sigma_k = fq-1\) in the symmetric case, and \(-1\) in the antisymmetric case

\begin{equation}
\rho(h) = - \frac{\psi(\Delta)}{\psi(1-\Delta)}\frac{\psi(1-\Delta - \frac{h}{2})}{\psi(\Delta - \frac{h}{2})}
\end{equation}

\(\psi(\Delta) = 2i \cos(\pi\Delta)\Gamma(1-2\Delta)\), and plugging this relation, we find that

\begin{equation}
g_k(h)=-\frac{1}{2}\frac{\tan\Big(\frac{\pi}{2}(h-1/2)\Big)}{h-1/2}\sigma_k
\end{equation}

therefore, the symmetric spectra is

\begin{equation}
g_{sym}(h)=-\frac{3}{2}\frac{\tan\Big(\frac{\pi}{2}(h-1/2)\Big)}{h-1/2}
\end{equation}

and the antisymmetric spectra

\begin{equation}
g_{anti}(h)=\frac{1}{2}\frac{\tan\Big(\frac{\pi}{2}(h-1/2)\Big)}{h-1/2}
\end{equation}

Now let us compare these results to ours. At $\beta = -\frac{1}{3}$, the spectra are as follows

\begin{equation}
\begin{split}
\MoveEqLeft
\qquad g_1(h) = -\frac{3}{2}\frac{\tan\Big(\frac{\pi}{2}(h-1/2)\Big)}{h-1/2} \\&
g_2(h) = -\frac{\tan\Big(\frac{\pi}{2}(h-1/2)\Big)}{h-1/2} \\&
g_3(h) = \frac{\tan\Big(\frac{\pi}{2}(h+1/2)\Big)}{h-1/2} \\&
g_4(h) = \frac{1}{2}\frac{\tan\Big(\frac{\pi}{2}(h-1/2)\Big)}{h-1/2}
\end{split}
\end{equation}

On the other hand

\begin{equation}
\begin{split}
\MoveEqLeft
O_1^{2n+1} = \psi_1 \partial_t^{2n+1} \psi_1 + \psi_2 \partial_t^{2n+1} \psi_2 \\&\qquad = \tilde\psi_1\partial_t^{2n+1} \tilde\psi_1 + \tilde\psi_2 \partial_t^{2n+1} \tilde\psi_2
\end{split}
\end{equation}
\begin{equation}
\begin{split}
\MoveEqLeft
O_4^{2n+1} = \psi_1 \partial_t^{2n+1} \psi_2 + \psi_2 \partial_t^{2n+1} \psi_1 \\&\qquad = \tilde\psi_1\partial_t^{2n+1} \tilde\psi_1 - \tilde\psi_2 \partial_t^{2n+1} \tilde\psi_2
\end{split}
\end{equation}

Hence, \(g_1\) and \(g_4\) corresponds to the operators studied in \cite{Gross:2016kjj}. The results clearly match, with \(g_1 = g_{sym}\), and \(g_4 = g_{anti}\). Note that \cite{Gross:2016kjj} does not include operators of the form \(\psi_1 \partial_t^{n} \psi_2\). Therefore, \(g_2\) and \(g_3\) cannot be compared to \cite{Gross:2016kjj}.

\subsection{Dualities in the Two Flavour Tensor Model}
As was shown previously, a duality between different values of \(\beta\) exists in this model. For example, we have just seen that \(\beta = 1\) is equivalent to \(\beta = 0\). This duality is due to the fact that the SU(2,R) transformation of Majoranas take us to a different basis in which the Hamiltonian is represented with a different coefficient for the non-coupling term.

Let us transform the Majorana fermions by \(\psi_1 = \frac{1}{\sqrt{2}}(\tilde\psi_1 + \tilde\psi_2)\), \(\psi_2 = \frac{1}{\sqrt{2}}(\tilde\psi_1 - \tilde\psi_2)\). The \(\tilde\psi\) clearly satisfies the fermion anti-commutation relations.

In doing the transformation, the term \(\psi_1^{a_1b_1c_1}\psi_1^{a_1b_2c_2}\psi_1^{a_2b_1c_2}\psi_1^{a_2b_2c_1}\) transforms as the following

\begin{equation}
\begin{split}
\MoveEqLeft
\psi_1^{a_1b_1c_1}\psi_1^{a_1b_2c_2}\psi_1^{a_2b_1c_2}\psi_1^{a_2b_2c_1} \\& = \frac{1}{4}(\tilde\psi_1^{a_1b_1c_1}+\tilde\psi_2^{a_1b_1c_1})(\tilde\psi_1^{a_1b_2c_2}+\tilde\psi_2^{a_1b_2c_2})(\tilde\psi_1^{a_2b_1c_2}+\tilde\psi_2^{a_2b_1c_2})(\tilde\psi_1^{a_2b_2c_1}+\tilde\psi_2^{a_2b_2c_1}) \\&
= \frac{1}{4}(\tilde\psi_1^{a_1b_1c_1}\tilde\psi_1^{a_1b_2c_2}\tilde\psi_1^{a_2b_1c_2}\tilde\psi_1^{a_2b_2c_1} + \tilde\psi_2^{a_1b_1c_1}\tilde\psi_2^{a_1b_2c_2}\tilde\psi_2^{a_2b_1c_2}\tilde\psi_2^{a_2b_2c_1}) \\& \qquad + \tilde\psi_1^{a_1b_1c_1}\tilde\psi_2^{a_1b_2c_2}\tilde\psi_2^{a_2b_1c_2}\tilde\psi_2^{a_2b_2c_1} + \tilde\psi_2^{a_1b_1c_1}\tilde\psi_1^{a_1b_2c_2}\tilde\psi_1^{a_2b_1c_2}\tilde\psi_1^{a_2b_2c_1} \\& \qquad + \frac{1}{2}(\tilde\psi_1^{a_1b_1c_1}\tilde\psi_1^{a_1b_2c_2}\tilde\psi_2^{a_2b_1c_2}\tilde\psi_2^{a_2b_2c_1} + \tilde\psi_1^{a_1b_1c_1}\tilde\psi_2^{a_1b_2c_2}\tilde\psi_1^{a_2b_1c_2}\tilde\psi_2^{a_2b_2c_1} + \tilde\psi_1^{a_1b_1c_1}\tilde\psi_2^{a_1b_2c_2}\tilde\psi_2^{a_2b_1c_2}\tilde\psi_1^{a_2b_2c_1})
\end{split}
\end{equation}

Similarly, \(\psi_2^{a_1b_1c_1}\psi_2^{a_1b_2c_2}\psi_2^{a_2b_1c_2}\psi_2^{a_2b_2c_1}\) transforms as

\begin{equation}
\begin{split}
\MoveEqLeft
\psi_2^{a_1b_1c_1}\psi_2^{a_1b_2c_2}\psi_2^{a_2b_1c_2}\psi_2^{a_2b_2c_1} \\&
= \frac{1}{4}(\tilde\psi_1^{a_1b_1c_1}\tilde\psi_1^{a_1b_2c_2}\tilde\psi_1^{a_2b_1c_2}\tilde\psi_1^{a_2b_2c_1} + \tilde\psi_2^{a_1b_1c_1}\tilde\psi_2^{a_1b_2c_2}\tilde\psi_2^{a_2b_1c_2}\tilde\psi_2^{a_2b_2c_1}) \\& \qquad - \tilde\psi_1^{a_1b_1c_1}\tilde\psi_2^{a_1b_2c_2}\tilde\psi_2^{a_2b_1c_2}\tilde\psi_2^{a_2b_2c_1} - \tilde\psi_2^{a_1b_1c_1}\tilde\psi_1^{a_1b_2c_2}\tilde\psi_1^{a_2b_1c_2}\tilde\psi_1^{a_2b_2c_1} \\& \qquad + \frac{1}{2}(\tilde\psi_1^{a_1b_1c_1}\tilde\psi_1^{a_1b_2c_2}\tilde\psi_2^{a_2b_1c_2}\tilde\psi_2^{a_2b_2c_1} + \tilde\psi_1^{a_1b_1c_1}\tilde\psi_2^{a_1b_2c_2}\tilde\psi_1^{a_2b_1c_2}\tilde\psi_2^{a_2b_2c_1} + \tilde\psi_1^{a_1b_1c_1}\tilde\psi_2^{a_1b_2c_2}\tilde\psi_2^{a_2b_1c_2}\tilde\psi_1^{a_2b_2c_1})
\end{split}
\end{equation}

Summing the two equations up
\begin{equation}
\begin{split}
\MoveEqLeft
\psi_1^{a_1b_1c_1}\psi_1^{a_1b_2c_2}\psi_1^{a_2b_1c_2}\psi_1^{a_2b_2c_1} + \psi_2^{a_1b_1c_1}\psi_2^{a_1b_2c_2}\psi_2^{a_2b_1c_2}\psi_2^{a_2b_2c_1} \\&
= \frac{1}{2}(\tilde\psi_1^{a_1b_1c_1}\tilde\psi_1^{a_1b_2c_2}\tilde\psi_1^{a_2b_1c_2}\tilde\psi_1^{a_2b_2c_1} + \tilde\psi_2^{a_1b_1c_1}\tilde\psi_2^{a_1b_2c_2}\tilde\psi_2^{a_2b_1c_2}\tilde\psi_2^{a_2b_2c_1}) \\& \qquad + (\tilde\psi_1^{a_1b_1c_1}\tilde\psi_1^{a_1b_2c_2}\tilde\psi_2^{a_2b_1c_2}\tilde\psi_2^{a_2b_2c_1} + \tilde\psi_1^{a_1b_1c_1}\tilde\psi_2^{a_1b_2c_2}\tilde\psi_1^{a_2b_1c_2}\tilde\psi_2^{a_2b_2c_1} + \tilde\psi_1^{a_1b_1c_1}\tilde\psi_2^{a_1b_2c_2}\tilde\psi_2^{a_2b_1c_2}\tilde\psi_1^{a_2b_2c_1})
\end{split}
\end{equation}



The other interaction term transforms in the following manner

\begin{equation}
\begin{split}
\MoveEqLeft
\psi_1^{a_1b_1c_1}\psi_1^{a_1b_2c_2}\psi_2^{a_2b_1c_2}\psi_2^{a_2b_2c_1} - \psi_1^{a_1b_1c_1}\psi_2^{a_1b_2c_2}\psi_1^{a_2b_1c_2}\psi_2^{a_2b_2c_1} + \psi_1^{a_1b_1c_1}\psi_2^{a_1b_2c_2}\psi_2^{a_2b_1c_2}\psi_1^{a_2b_2c_1} \\&
= \frac{3}{4}(\tilde\psi_1^{a_1b_1c_1}\tilde\psi_1^{a_1b_2c_2}\tilde\psi_1^{a_2b_1c_2}\tilde\psi_1^{a_2b_2c_1} + \tilde\psi_2^{a_1b_1c_1}\tilde\psi_2^{a_1b_2c_2}\tilde\psi_2^{a_2b_1c_2}\tilde\psi_2^{a_2b_2c_1})\\& \qquad - \frac{1}{2}(\tilde\psi_1^{a_1b_1c_1}\tilde\psi_1^{a_1b_2c_2}\tilde\psi_2^{a_2b_1c_2}\tilde\psi_2^{a_2b_2c_1} - \tilde\psi_1^{a_1b_1c_1}\tilde\psi_2^{a_1b_2c_2}\tilde\psi_1^{a_2b_1c_2}\tilde\psi_2^{a_2b_2c_1} + \tilde\psi_1^{a_1b_1c_1}\tilde\psi_2^{a_1b_2c_2}\tilde\psi_2^{a_2b_1c_2}\tilde\psi_1^{a_2b_2c_1})
\end{split}
\end{equation}

Hence, adding the two interaction terms up, one finds the following correspondence between Hamiltonians with different \(\beta\)s

\begin{equation}
\begin{split}
\MoveEqLeft
H = \frac{\beta g}{2}(\psi_1^{a_1b_1c_1}\psi_1^{a_1b_2c_2}\psi_2^{a_2b_1c_2}\psi_2^{a_2b_2c_1} - \psi_1^{a_1b_1c_1}\psi_2^{a_1b_2c_2}\psi_1^{a_2b_1c_2}\psi_2^{a_2b_2c_1} + \psi_1^{a_1b_1c_1}\psi_2^{a_1b_2c_2}\psi_2^{a_2b_1c_2}\psi_1^{a_2b_2c_1}) \\&
+ \frac{g}{4}(\psi_1^{a_1b_1c_1}\psi_1^{a_1b_2c_2}\psi_1^{a_2b_1c_2}\psi_1^{a_2b_2c_1}+\psi_2^{a_1b_1c_1}\psi_2^{a_1b_2c_2}\psi_2^{a_2b_1c_2}\psi_2^{a_2b_2c_1}) \\&
= \frac{(3\beta+1)g}{8}(\tilde\psi_1^{a_1b_1c_1}\tilde\psi_1^{a_1b_2c_2}\tilde\psi_1^{a_2b_1c_2}\tilde\psi_1^{a_2b_2c_1} + \tilde\psi_2^{a_1b_1c_1}\tilde\psi_2^{a_1b_2c_2}\tilde\psi_2^{a_2b_1c_2}\tilde\psi_2^{a_2b_2c_1})\\& \quad + \frac{(-\beta+1)g}{4}(\tilde\psi_1^{a_1b_1c_1}\tilde\psi_1^{a_1b_2c_2}\tilde\psi_2^{a_2b_1c_2}\tilde\psi_2^{a_2b_2c_1} - \tilde\psi_1^{a_1b_1c_1}\tilde\psi_2^{a_1b_2c_2}\tilde\psi_1^{a_2b_1c_2}\tilde\psi_2^{a_2b_2c_1} + \tilde\psi_1^{a_1b_1c_1}\tilde\psi_2^{a_1b_2c_2}\tilde\psi_2^{a_2b_1c_2}\tilde\psi_1^{a_2b_2c_1}) \\&
= \frac{\beta' g'}{2}(\tilde\psi_1^{a_1b_1c_1}\tilde\psi_1^{a_1b_2c_2}\tilde\psi_2^{a_2b_1c_2}\tilde\psi_2^{a_2b_2c_1} - \tilde\psi_1^{a_1b_1c_1}\tilde\psi_2^{a_1b_2c_2}\tilde\psi_1^{a_2b_1c_2}\tilde\psi_2^{a_2b_2c_1} + \tilde\psi_1^{a_1b_1c_1}\tilde\psi_2^{a_1b_2c_2}\tilde\psi_2^{a_2b_1c_2}\tilde\psi_1^{a_2b_2c_1}) \\& \qquad + \frac{g'}{4}(\tilde\psi_1^{a_1b_1c_1}\tilde\psi_1^{a_1b_2c_2}\tilde\psi_1^{a_2b_1c_2}\tilde\psi_1^{a_2b_2c_1} + \tilde\psi_2^{a_1b_1c_1}\tilde\psi_2^{a_1b_2c_2}\tilde\psi_2^{a_2b_1c_2}\tilde\psi_2^{a_2b_2c_1})
\end{split}
\end{equation}

Here

\begin{equation}
g' = \frac{(3\beta+1)g}{2}, \quad \quad \beta' = \frac{-\beta+1}{3\beta+1}
\end{equation}

Therefore, \(H_\beta\) and \(H_{\beta'}\) where \(\beta' = \frac{-\beta+1}{3\beta+1}\) are equivalent up to a scaling of the interaction strength. This equivalence relation is reflected in the behavior of \(g_i(h)\).

\(g_1(h)\) is naturally invariant under \(\beta \rightarrow \beta'\). Now let us look at how \(g_2(h)\) transforms

\begin{equation}
\begin{split}
\MoveEqLeft
g'_2(h) = -\frac{3}{2}\frac{-\beta'^2+1}{3\beta'^2+1}\frac{\tan(\frac{\pi}{2}(h-1/2))}{h-1/2} \\& \qquad
= -\frac{3}{2}\frac{-(-\beta+1)^2+(3\beta+1)^2}{3(-\beta+1)^2+(3\beta+1)^2}\frac{\tan(\frac{\pi}{2}(h-1/2))}{h-1/2} \\& \qquad
= -\frac{3}{2}\frac{8\beta^2+8\beta}{12\beta^2+4}\frac{\tan(\frac{\pi}{2}(h-1/2))}{h-1/2} \\& \qquad
= g_4(h)
\end{split}
\end{equation}

Hence, \(g_2(h)\) transforms to \(g_4(h)\). Now let us looks at \(g_3(h)\)

\begin{equation}
\begin{split}
\MoveEqLeft
g_3(h) = \frac{3\beta'^2-3\beta'}{3\beta'^2+1}\frac{\tan(\frac{\pi}{2}(h+1/2))}{h-1/2} \\& \qquad
= \frac{3(-\beta+1)^2-3(-\beta+1)(3\beta+1)}{3(-\beta+1)^2+(3\beta+1)^2}\frac{\tan(\frac{\pi}{2}(h+1/2))}{h-1/2} \\& \qquad
= \frac{12\beta^2-12\beta}{12\beta^2+4}\frac{\tan(\frac{\pi}{2}(h+1/2))}{h-1/2} \\& \qquad
= \frac{3\beta^2-3\beta}{3\beta^2+1}\frac{\tan(\frac{\pi}{2}(h+1/2))}{h-1/2}
\end{split}
\end{equation}

Hence, \(g_3(h)\) transforms to itself.

All this behavior makes sense: with the \(\psi\) to \(\tilde\psi\) transformation, \(O_i\) operators transform as the following

\begin{equation}
\begin{split}
\MoveEqLeft
O_1 = \psi_1 \partial^{2n+1}\psi_1+\psi_2 \partial^{2n+1}\psi_2 \\& \qquad
= \tilde\psi_1 \partial^{2n+1} \tilde\psi_1 + \tilde\psi_2 \partial^{2n+1} \tilde\psi_2
= O'_1
\end{split}
\end{equation}

\begin{equation}
\begin{split}
\MoveEqLeft
O_2 = \psi_1 \partial^{2n+1}\psi_1-\psi_2 \partial^{2n+1}\psi_2 \\& \qquad
= \tilde\psi_1 \partial^{2n+1} \tilde\psi_2 + \tilde\psi_2 \partial^{2n+1} \tilde\psi_1
= O'_4
\end{split}
\end{equation}

\begin{equation}
\begin{split}
\MoveEqLeft
O_3 = \psi_1 \partial^{2n}\psi_2-\psi_2 \partial^{2n+1}\psi_1 \\& \qquad
= \tilde\psi_2 \partial^{2n+1} \tilde\psi_1 - \tilde\psi_1 \partial^{2n+1} \tilde\psi_2
= O'_3
\end{split}
\end{equation}

The way that \(O_1\) and \(O_4\) transform to itself (upto a sign for \(O_4\)), and the way that \(O_2\) transform to \(O_3\) and vice versa provides a clear prediction to the behavior of \(g_i(h)\) under the duality transformations.

Note that this duality is not a perfect one: as \(g\) scales with the transformation, the energy spectrum also scales at the same time, although the relative ratios stay the same. The operator dimensions stay the same, however, because the normalization of \(g\) cancels in them.

Also, $g'$ goes to 0 as \(\beta \rightarrow -\frac{1}{3}\). This is due to the definition of the Hamiltonian. If we were to define $\tilde g = \alpha g$, and $\alpha = 1/\beta$, then the transformation would be

\begin{equation}
\tilde g' = \frac{\alpha-1}{2}\tilde g \qquad \& \qquad \beta' = \frac{\alpha+3}{\alpha-1}
\end{equation}

With this definition, $\tilde g'$ does not go to zero. However, $\tilde g'$ still goes to zero as \(\alpha\) approaches $1$, which is problematic in its own way.

\subsection{Complex Scaling Dimensions}
In this section, we examine if there exist any complex roots that satisfy the equation \(g(h) = 1\). If such complex roots were to exist, then that means that a conformal primary has a complex scaling dimension, which leads to a destabilization of the model.

Let us write \(h - 1/2 = x\). Then, we need to have the following, for the coefficients in \(g_i\) infront of the part dependent on \(h\) are all real numbers

\begin{equation}
Im \Bigg\{\frac{\tan \Big(\frac{\pi}{2}x\Big)}{x}\Bigg\} = 0
\end{equation}

for \(g_1\), \(g_2\), and \(g_4\). For \(g_3\), we need to have 

\begin{equation}
Im \Bigg\{\frac{\cot \Big(\frac{\pi}{2}x\Big)}{x}\Bigg\} = 0
\end{equation}

Now, writing \(x = z_1 + iz_2\), where \(z_1, z_2 \in R\) we find the following

\begin{equation}
\frac{\tan \Big(\frac{\pi}{2}x\Big)}{x} = \frac{\frac{-\sinh \frac{\pi}{2}z_2 \cosh \frac{\pi}{2}z_2 + i \sin \frac{\pi}{2}z_1 \cos \frac{\pi}{2}z_1}{\sinh^2\frac{\pi}{2}z_2 - \cos^2\frac{\pi}{2}x}}{-z_2+iz_1}
\end{equation}
\begin{equation}
\frac{\cot \Big(\frac{\pi}{2}x\Big)}{x} = \frac{\frac{-\sinh \frac{\pi}{2}z_2 \cosh \frac{\pi}{2}z_2 - i \sin \frac{\pi}{2}z_1 \cos \frac{\pi}{2}z_1}{\sinh^2\frac{\pi}{2}z_2 - \sin^2\frac{\pi}{2}x}}{-z_2+iz_1}
\end{equation}

Since \(\sinh^2\frac{\pi}{2}z_2 - \sin^2\frac{\pi}{2}x\) is real

\begin{equation}
-\sinh(\pi z_2) / \sin(\pi z_1) = -z_2 / z_1
\end{equation}
\begin{equation}
\sinh(\pi z_2) / \sin(\pi z_1) = -z_2 / z_1
\end{equation}

we find that the only solutions to the previous equation with \(z_2 \neq 0\) happens at \(z_1 = 0\). Therefore, complex roots of \(g_i(h) = 1\) take the form \(\frac{1}{2} + ik\). Now, plugging this in, we find the following

\begin{equation}
g_1\Big(\frac{1}{2} + ik\Big) = -\frac{3}{2}\frac{\tanh\Big(\frac{\pi}{2}k\Big)}{k}
\end{equation}
\begin{equation}
g_2\Big(\frac{1}{2} + ik\Big) = -\frac{3}{2}\frac{-\beta^2+1}{3\beta^2+1}\frac{\tanh\Big(\frac{\pi}{2}k\Big)}{k}
\end{equation}
\begin{equation}
g_3\Big(\frac{1}{2} + ik\Big) = \frac{3\beta^2-3\beta}{3\beta^2+1}\frac{\coth\Big(\frac{\pi}{2}k\Big)}{k}
\end{equation}
\begin{equation}
g_4\Big(\frac{1}{2} + ik\Big) = -\frac{3\beta^2+3\beta}{3\beta^2+1}\frac{\tanh\Big(\frac{\pi}{2}k\Big)}{k}
\end{equation}

\(\frac{\tanh\frac{\pi}{2}k}{k}\) moves between 0 and \(\pi/2\). Hence, with some algebra, we find that \(g_1\), \(g_2\), \(g_4\) never can equal 1 where \(h = \frac{1}{2} + ik\). Consequently, they have no complex roots.

\begin{figure}
  \begin{center}
    \includegraphics [width=0.45\textwidth, angle=0.]{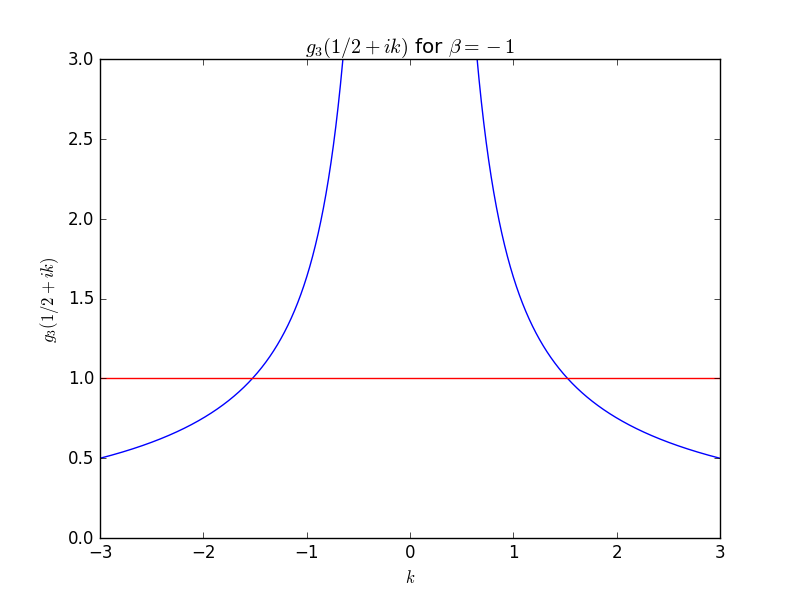}
    \includegraphics [width=0.45\textwidth, angle=0.]{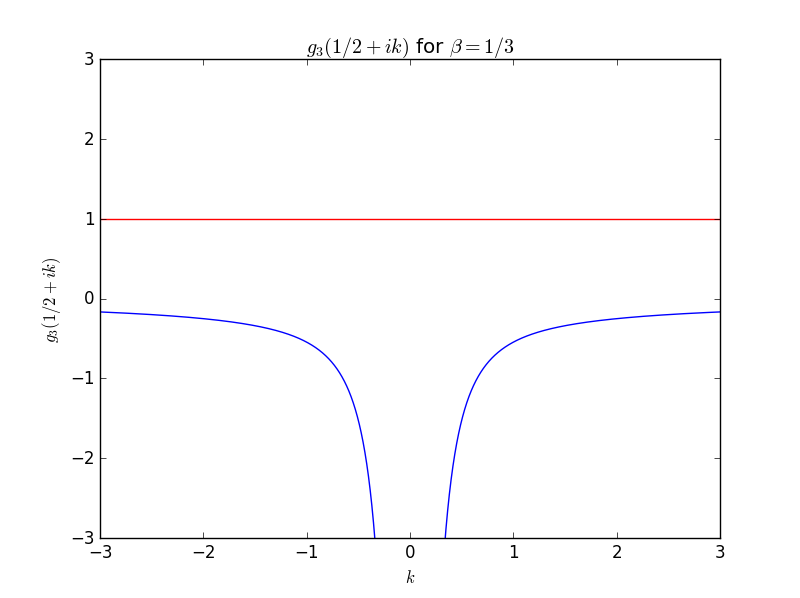}    
  \end{center}
  \caption{Left: \(g_3 \big(\frac{1}{2}+ik \big)\) at \(\beta = -1\), Right: \(g_3\big(\frac{1}{2}+ik \big)\) at \(\beta = \frac{1}{3}\)}
  \label{gcomp}
\end{figure}

On the other hand, \(\frac{\coth\frac{\pi}{2}k}{k}\) moves between 0 and $\infty$. Therefore, $g_3$ has roots of the form \(\frac{1}{2}+ik\) when \(\frac{3\beta^2-3\beta}{3\beta^2+1}\) is larger than 0. Therefore, when \(\beta > 1\) or \(\beta < 0\) $g_3$ has complex roots. This point is illustrated in figure \ref{gcomp}. In the left figure of $\beta = -1$, it is evident that there exist roots of the form \(\frac{1}{2}+ik\), whereas for the right figure of $\beta = \frac{1}{3}$, they do not.

Consequently, for \(\beta > 1\), since the conformal primary \(\psi_1 \partial_t^{2n}\psi_2 - \psi_2\partial_t^{2n}\psi_1\) has a complex scaling dimension, the model is rendered unstable. Going back to section 2.2.2, the reason why the real scaling dimension near 0.5 disappears as $\beta$ crosses 1 from below is because of this instability.

\section{Conclusion}
In this thesis, we have investigated the two flavour tensor model with the Hamiltonian in equation \ref{Hamiltonian}. This two flavour tensor model exhibits melonic dominance in the large \(N\) limit. Using the Schwinger - Dyson equation, we found the propagator and determined it to be the following at large interactions

\begin{equation}
\begin{split}
G(t_2 - t_1) = -\left(\frac{1}{4\pi(3\beta^2+1)g^2N^3}\right)^{\frac{1}{4}}\frac{\sgn(t_2-t_1)}{|t_2-t_1|^{1/2}}
\end{split}
\end{equation}

We went on to compute the kernels of the four point functions \(\braket{\psi_1^{abc}\psi_1^{abc}\psi_1^{a'b'c'}\psi_1^{a'b'c'}}\), and \(\braket{\psi_1^{abc}\psi_1^{abc}\psi_2^{a'b'c'}\psi_2^{a'b'c'}}\). Using these four point functions we determined the spectrum of the operators \(O^{2n+1}_{1,2} = \psi_1 \partial_t^{2n+1} \psi_1 \pm \psi_2 \partial_t^{2n+1} \psi_2\) to be to be the following

\begin{equation}
g_1(h) = -\frac{3}{2}\frac{\tan\Big((\frac{\pi}{2}(h-\frac{1}{2})\Big)}{h-\frac{1}{2}}
\end{equation}
\begin{equation}
g_2(h) = -\frac{3}{2}\frac{-\beta^2+1}{3\beta^2+1}\frac{\tan\Big((\frac{\pi}{2}(h-\frac{1}{2})\Big)}{h-\frac{1}{2}}
\end{equation}

Similarly, we calculated the kernels of the four point functions \(\braket{\psi_1^{abc}\psi_2^{abc}\psi_1^{a'b'c'}\psi_2^{a'b'c'}}\), and \(\braket{\psi_1^{abc}\psi_2^{abc}\psi_2^{a'b'c'}\psi_1^{a'b'c'}}\). With it, we determined the spectrum of the operators \(O^{2n}_{3} = \psi_1 \partial_t^{2n} \psi_2 - \psi_2 \partial_t^{2n} \psi_1\) and \(O^{2n+1}_{4} = \psi_1 \partial_t^{2n+1} \psi_2 + \psi_2 \partial_t^{2n+1} \psi_1\) to be the following

\begin{equation}
g_3(h) = \frac{3\beta^2-3\beta}{3\beta^2+1}\frac{\tan\Big((\frac{\pi}{2}(h+\frac{1}{2})\Big)}{h-\frac{1}{2}}
\end{equation}
\begin{equation}
g_4(h) = -\frac{3\beta^2+3\beta}{3\beta^2+1}\frac{\tan\Big((\frac{\pi}{2}(h+\frac{1}{2})\Big)}{h-\frac{1}{2}}
\end{equation}

We compared the results of the spectra that we obtained here with the two flavour SYK model that Gross and Rosenhaus studied \cite{Gross:2016kjj}, and the flavourless tensor model and the complex bipartite tensor model studied by Klebanov and Tarnopolsky \cite{Klebanov:2016xxf, Klebanov:2018fzb}. Both show a good match.

Using the spectra of the operators \(O\), it is possible to find their scaling dimensions by finding the roots of \(g_i(h) = 1\). A special case that we looked at is when \(\beta = 1\). We determined the scaling dimensions of \(O_1\) and \(O_4\) to be 2.00, 3.77, 5.68, 7.63, \(\dots\) and converging to \(2n + 3/2\). For \(O_2\), we found the scaling dimensions to be \(2n+3/2\), and for \(O_3\), to be \(2n+1/2\).

At \(\beta = 1\), we found that a rotation of the Majorana fermions decouples the Hamiltonian into two separate \(O(N)^3\). This decoupling explains why the scaling dimensions of \(O_1\) and \(O_4\) are even or odd integers plus \(1/2\), and suggest that the operators \(\psi_1 \partial_t^n \psi_2\) is going to be of dimension \(n+1/2\).

At \(\beta = -\frac{1}{3}\), we found that with a rotation of the Majorana fermions, the model becomes equivalent to the two flavour SYK model studied by Gross and Rosenhaus in \cite{Gross:2016kjj}. We checked that the spectra of the \(\psi_i \partial_t^{2n+1} \psi_i\) is identical to one another.

In addition we studied the duality relations between two different values of $\beta$. Rotating the Majorana fermions by 45 degrees, we arrived at the conclusion that the theory with the coefficient \(\beta\) in the Hamiltonian is equivalent to that with

\begin{equation}
g' = \frac{(3\beta+1)g}{2}, \quad \quad \beta' = \frac{-\beta+1}{3\beta+1}
\end{equation}

This rotation transform \(g_2\) to \(g_4\) and vice versa, and \(g_1\) and \(g_3\) to themselves. This is not a true duality, for \(g' = (3\beta+1)g/2\), and hence the energy levels scale too. Nevertheless, the ratio of the energy eigenvalues is preserved. Furthermore, the operator dimensions are preserved for the normalization of \(g\) cancels in the scaling dimensions.

Last, we looked for the existence of complex roots to the equation \(g(h) = 1\). We found that for \(\beta > 1\) or \(\beta < 0\), there always exist a pair of complex roots of the form $\frac{1}{2}+ik$, and hence we conclude that the model is stable only for  \(0< \beta < 1\).

In turn, predictions can be made on the variables of the holographic gravity dual. Formal energy levels of the form $\frac{1}{2}+ik$ correspond to scalar fields in \(AdS_2\) whose \(m^2\) is below the Breitenlohner-Freedman bound \(m_{BF}^2 = -\frac{1}{4}\). Since $\Delta = \frac{1}{2} \pm \sqrt{\frac{1}{4}+m^2}$ \cite{Maldacena:1997re, Gubser:1998bc, Witten:1998qj}, we predict

\begin{equation}
m^2 = -\frac{1}{4} - k^2
\end{equation}

Where \(k\) is a root of \(g_3\big(\frac{1}{2}+ik\big)=1\). For the complex bipartite model, it is found that \(m^2 = -2.576\), and for the two flavour SYK model,  \(m^2 = -1.398\).

\newpage
\appendix
\section{Graphical Notation of the Hamiltonian}
The calculations in this thesis were greatly simplified with the use of graphical notations. For example, in the calculation of the kernel \(K_{reg}\), one had to contract hundreds of delta functions with one another, leading to computations that spanned several pages. However, using graphical notations, this calculation can be done with simple comparisons of tetrahedrons.

\begin{figure}
\begin{center}
\includegraphics[scale=0.4]{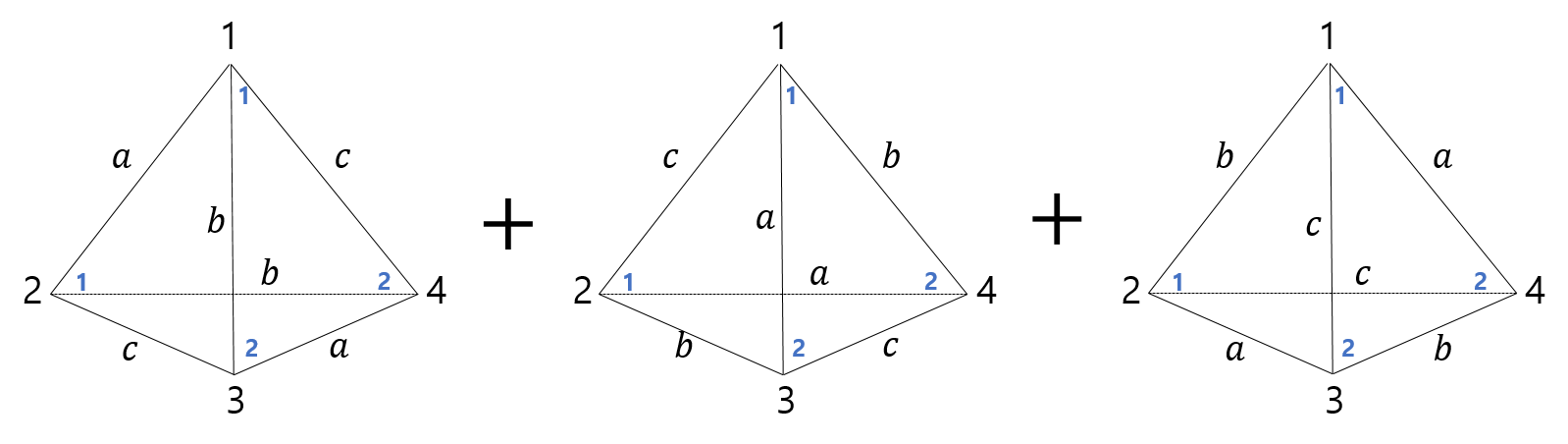}
\end{center}
\caption{The first term of the Hamiltonian, \(\psi_1^{a_1b_1c_1}\psi_1^{a_1b_2c_2}\psi_2^{a_2b_1c_2}\psi_2^{a_2b_2c_1}-\psi_1^{a_1b_1c_1}\psi_1^{a_2b_1c_2}\psi_2^{a_1b_2c_2}\psi_2^{a_2b_2c_1}+\psi_1^{a_1b_1c_1}\psi_1^{a_2b_1c_2}\psi_2^{a_2b_2c_1}\psi_2^{a_1b_2c_2}\) in graphical notations. Each vertex is a Majorana, whose order is denoted by the black number. The blue numbers denote the flavour. Each edge is a delta function between the indices of the Majoranas.}
\label{ftham}
\end{figure}

\begin{figure}
\begin{center}
\includegraphics[scale=0.4]{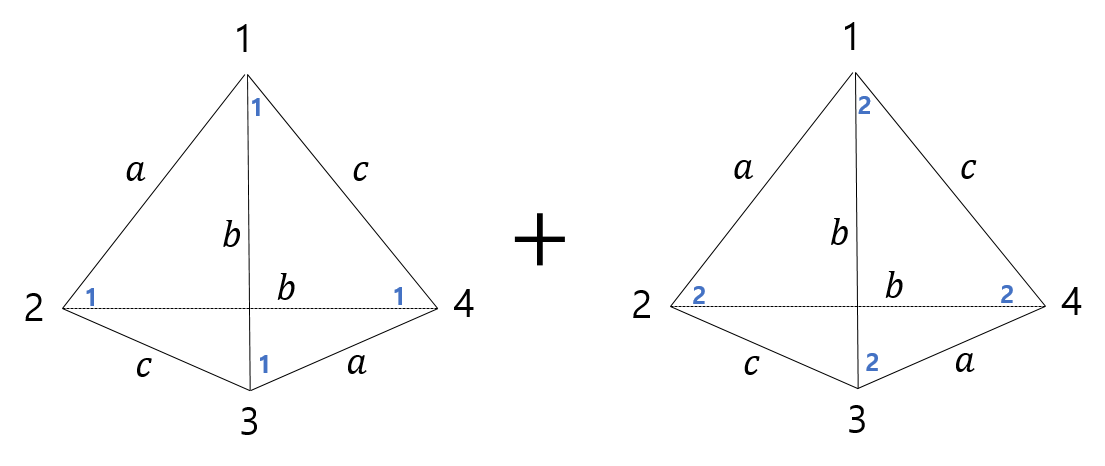}
\end{center}
\caption{The Second Term of the Hamiltonian, \(\psi_1^{a_1b_1c_1}\psi_1^{a_1b_2c_2}\psi_1^{a_2b_1c_2}\psi_1^{a_2b_2c_1}\) in Graphical Notations. Each vertex is a Majorana, whose order is denoted by the black number. The blue numbers denote the flavour. Each edge is a delta function between the indices of the Majoranas.}
\label{stham}
\end{figure}

In short, each \(\psi^4\) term in the Hamiltonian can be represented with a tetrahedron as illustrated in figure \ref{ftham} and \ref{stham}. This is done by writing the Hamiltonian with several delta functions. Recall that the Hamiltonian was written as the following

\begin{equation}
\begin{split}
\MoveEqLeft
H = \frac{\beta g}{2} I_1(a_i, b_i, c_i) \psi_1^{a_1 b_1 c_1} \psi_1^{a_2 b_2 c_2} \psi_2^{a_3 b_3 c_3} \psi_2^{a_4 b_4 c_4} \\& \qquad\qquad + \frac{g}{4}I_2(a_i, b_i, c_i)(\psi_1^{a_1 b_1 c_1}\psi_1^{a_2 b_2 c_2} \psi_1^{a_3 b_3 c_3} \psi_1^{a_4 b_4 c_4}+\psi_2^{a_1 b_1 c_1}\psi_2^{a_2 b_2 c_2} \psi_2^{a_3 b_3 c_3} \psi_2^{a_4 b_4 c_4})
\end{split}
\end{equation}

where
\begin{equation}
I_1(a_i, b_i, c_i) = \delta_{a_1 a_2}\delta_{b_1 b_3}\delta_{c_1 c_4}\delta_{b_2 b_4}\delta_{c_2 c_3}\delta_{a_3 a_4} - \delta_{a_1 a_3}\delta_{b_1 b_2}\delta_{c_1 c_4}\delta_{a_2 a_4}\delta_{c_2 c_3}\delta_{b_3 b_4} + \delta_{a_1 a_3}\delta_{b_1 b_4}\delta_{c_1 c_2}\delta_{a_2 a_4}\delta_{b_2 b_3}\delta_{c_3 c_4}
\end{equation}
\begin{equation}
I_2(a_i, b_i, c_i) = \delta_{a_1 a_2}\delta_{b_1 b_3}\delta_{c_1 c_4}\delta_{b_2 b_4}\delta_{c_2 c_3}\delta_{a_3 a_4}
\end{equation}

Now, in this expression, one can consider \(\psi_i^{a_jb_jc_j}\) as the vertex \(i\) of a tetrahedron, and each \(\delta\) function as an edge of a tetrahedron. For examples, compare the first term of \(I_1\) with the leftmost figure of \ref{ftham}.

Vertices 1 and 2, and 3 and 4 are connected in the \(a\) indice, which corresponds to \(\delta_{a_1a_2}\) and \(\delta_{a_3a_4}\). Similarly, vertices 1 and 3, and 2 and 4 are connected in the \(b\) color, and this leads to \(\delta_{b_1b_3}\) and \(\delta_{b_2b_4}\). Last, vertices 1 and 4, and 2 and 3 are connected in the c color, and this corresponds to \(\delta_{c_1c_4}\) and \(\delta_{c_2c_3}\). Multiplying all these delta functions, one arrives at the first term of \(I_1(a_i, b_i, c_i)\).

All in all, the first term in the Hamiltonian (minus the factor) corresponds to figure \ref{ftham}, and the second term, to figure \ref{stham}. Now let us use this graphical representation to simplify computations.

We shall calculate the first term of the kernel that we found for \(\braket{\psi_1^{abc}(t_1)\psi_1^{abc}(t_2)\psi_1^{abc}(t_3)\psi_1^{abc}(t_4}\) and \(\braket{\psi_1^{abc}(t_1)\psi_1^{abc}(t_2)\psi_2^{abc}(t_3)\psi_2^{abc}(t_4}\). The kernel is created from the wick contractions of

\begin{equation}
\frac{\beta^2g^2}{4}I_1(a_i,b_i,c_i)I_1(a'_i,b'_i,c'_i)\psi_1^{a_1b_1c_1}\psi_1^{a_2b_2c_2}\psi_2^{a_3b_3c_3}\psi_2^{a_4b_4c_4}\psi_1^{a'_1b'_1c'_1}\psi_1^{a'_2b'_2c'_2}\psi_2^{a'_3b'_3c'_3}\psi_2^{a'_4b'_4c'_4}\psi_2^{a'b'c'}\psi_2^{a'b'c'}
\end{equation}

\begin{figure}
\begin{center}
\includegraphics[scale=0.4]{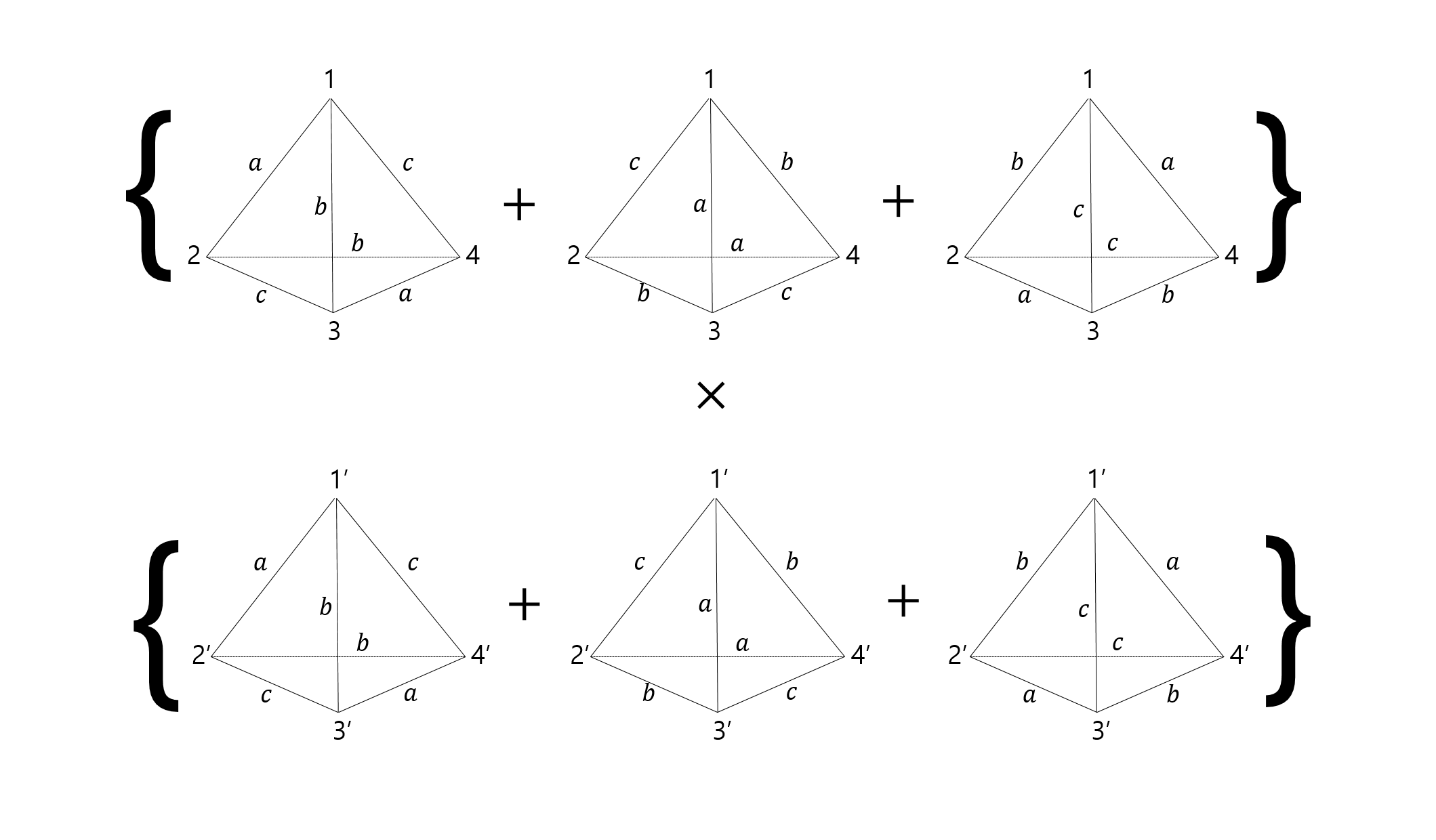}
\end{center}
\caption{\(I_1(a_i,b_i,c_i) \times I_1(a'_i,b'_i,c'_i)\) is of the form above in the diagrammatic representation.}
\label{kernelcalc}
\end{figure}

This object, written without \(\psi_2^{a'b'c'}\psi_2^{a'b'c'}\) and expressed graphically is as illustrated in figure \ref{kernelcalc}.

Now let us contract \(\psi_2^{a'b'c'}\)s with \(\psi_2^{a_4b_4c_4}\) and \(\psi_2^{a'_4b'_4c'_4}\). Since the Hamiltonian is symmetric under the interchange of  \(a_3b_3c_3\) with \(a_4b_4c_4\), it is enough to multiply the resulting object by 4 and not do further contractions of \(\psi_2^{a'b'c'}\) with other \(\psi_2\)s.

\begin{figure}
\begin{center}
\includegraphics[scale=0.4]{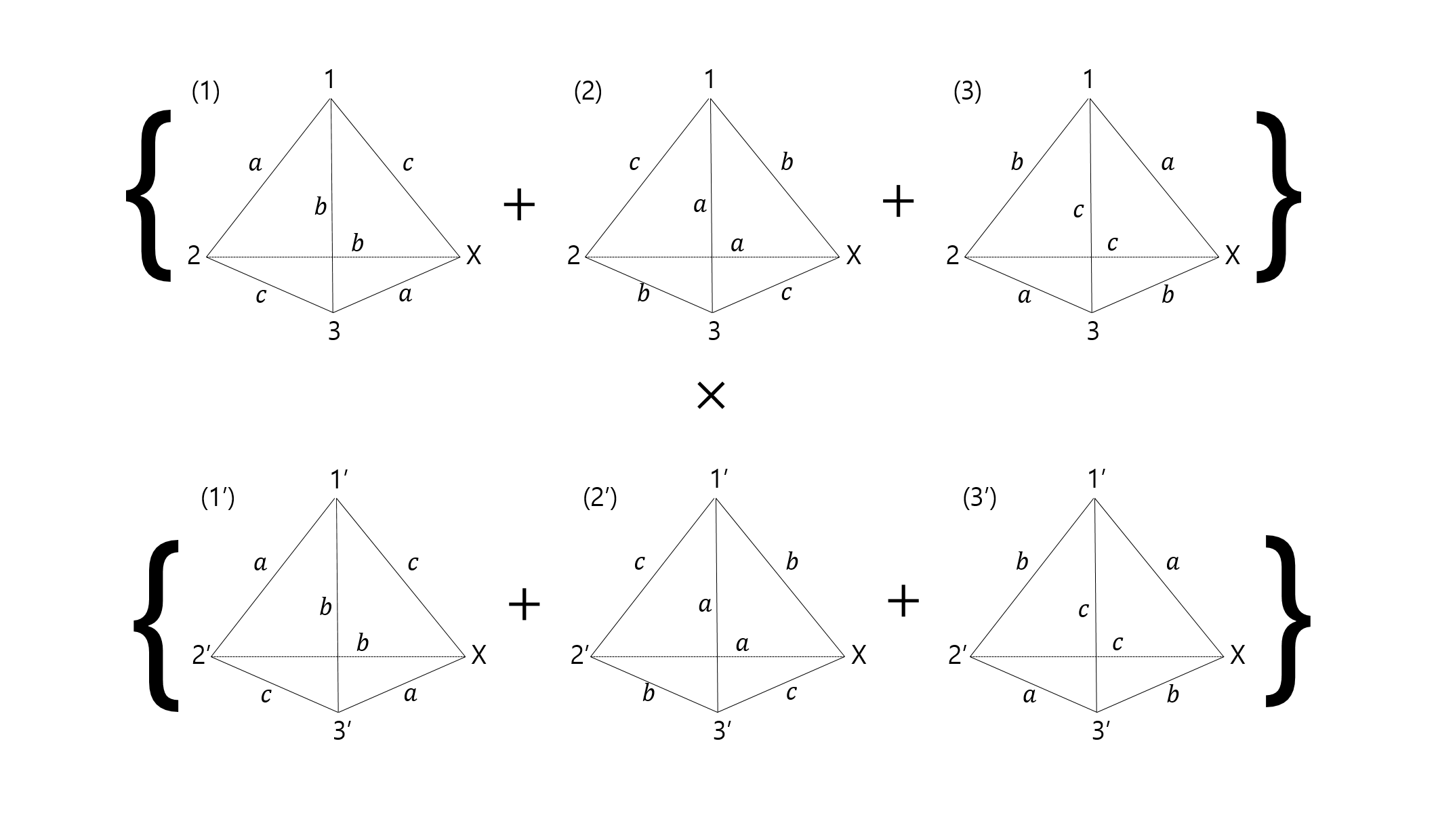}
\end{center}
\caption{The calculation of the kernel, after the contraction of \(\psi_2^{a'b'c'}\)s with \(\psi_2^{a_4b_4c_4}\) and \(\psi_2^{a'_4b'_4c'_4}\).}
\label{kernelcalc1}
\end{figure}

As a result of the contraction, one obtains \(\delta_{a_4a'}\delta_{b_4b'}\delta_{c_4c'}\delta_{a'_4a'}\delta_{b'_4b'}\delta_{c'_4c'}\). This leads to \(\delta_{aa'}\delta_{bb'}\delta_{cc'}\), and hence one can equate vertex 4 with vertex 4'. This change is reflected in figure \ref{kernelcalc1}.

Now, one can perform wick contractions between 1 and 2 with 1' and 2', and 3 with 3'. One must only choose two contractions from the three available contractions for two \(\psi_i\)s need to contract with the \(\psi_i\)s coming from the left of the kernel.

\begin{figure}
\begin{center}
\includegraphics[scale=0.4]{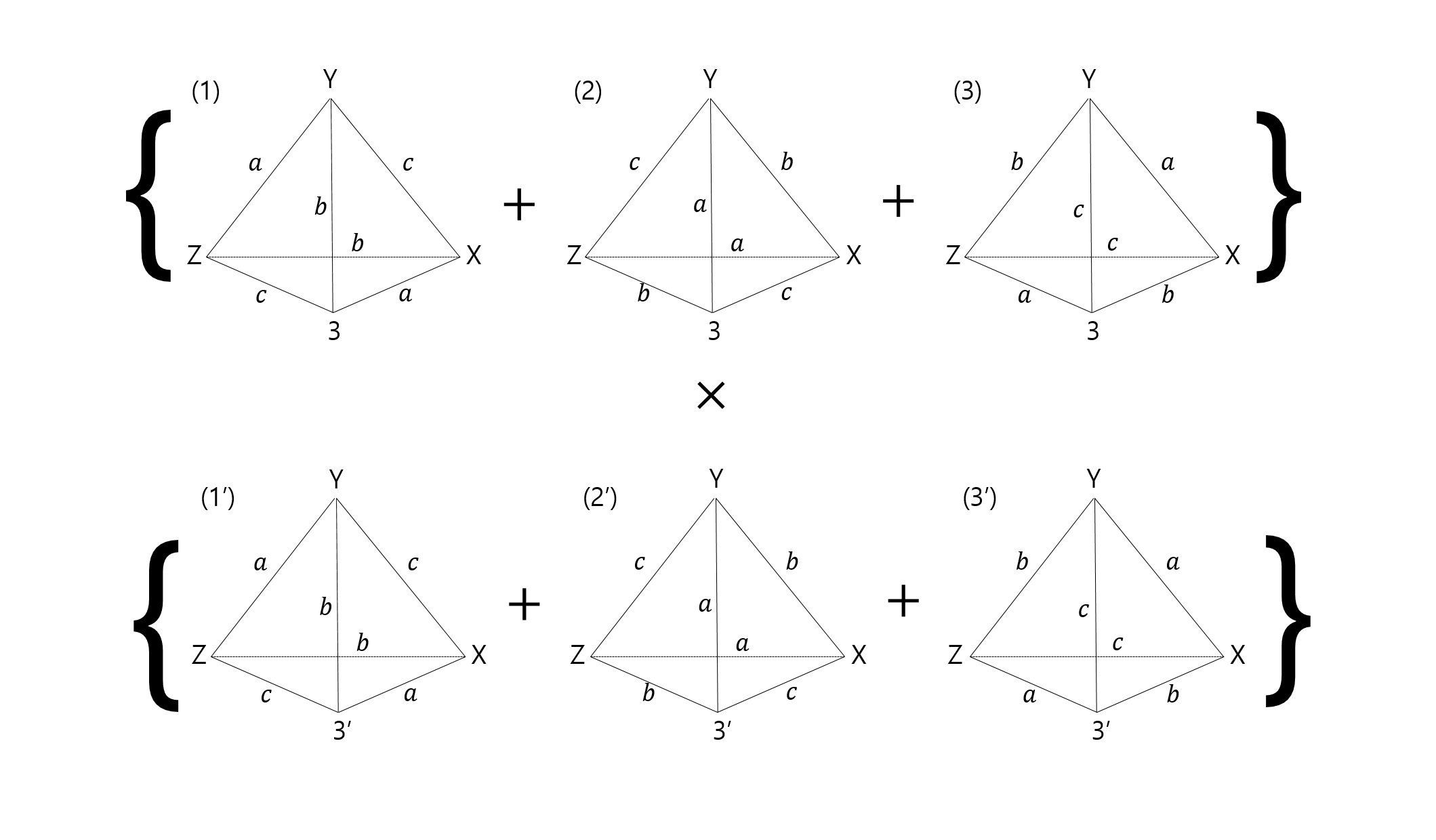}
\end{center}
\caption{The calculation of the kernel, after the contraction of \(\psi_2^{a_1b_1c_1}\) with \(\psi_2^{a'_1b'_1c'_1}\) and \(\psi_2^{a_2b_2c_2}\) with \(\psi_2^{a'_2b'_2c'_2}\).}
\label{kernelcalc2}
\end{figure}

Suppose we perform wick contractions between 1 and 1', and 2 and 2'. We obtain what is illustrated in figure \ref{kernelcalc2}.

Now lifting out the parentheses, we find 9 pairs of tetrahedrons. Note that one only needs to perform calculations on a choice of three pairs only, a viable choice of which would be tetrahedron pairs (1)(1'), (1)(2'), and (1)(3'). This is because the Hamiltonian is symmetric under the cyclic rotation of \(a\), \(b\), and \(c\): (2)(2'), (2)(3'), and (2)(1') would return exactly the same result, and the same goes for pairs that contain (3).

Now, pair (1)(1') gives

\begin{equation}
\begin{split}
\MoveEqLeft
\delta_{a_Ya_Z}\delta_{a_Xa_3}\delta_{b_Yb_3}\delta_{b_Zb_X}\delta_{c_Zc_3}\delta_{c_Xc_Y}\delta_{a_Ya_Z}\delta_{a_Xa_3'}\delta_{b_Yb_3'}\delta_{b_Zb_X}\delta_{c_Zc_3'}\delta_{c_Xc_Y}\psi_2^{a_3b_3c_3}\psi_2^{a'_3b'_3c'_3} \\&
\qquad = N^3 \delta_{a_3a'_3}\delta_{b_3b'_3}\delta_{c_3c'_3}\psi_2^{a_3b_3c_3}\psi_2^{a'_3b'_3c'_3}
\end{split}
\end{equation}

On the other hand, pair (1)(2') gives

\begin{equation}
\begin{split}
\MoveEqLeft
\delta_{a_Ya_Z}\delta_{a_Xa_3}\delta_{b_Yb_3}\delta_{b_Zb_X}\delta_{c_Zc_3}\delta_{c_Xc_Y}\delta_{c_Yc_Z}\delta_{c_Xc_3'}\delta_{a_Ya_3'}\delta_{a_Za_X}\delta_{b_Zb_3'}\delta_{b_Xb_Y}\psi_2^{a_3b_3c_3}\psi_2^{a'_3b'_3c'_3} \\&
\qquad = \delta_{a_3a'_3}\delta_{b_3b'_3}\delta_{c_3c'_3}\psi_2^{a_3b_3c_3}\psi_2^{a'_3b'_3c'_3}
\end{split}
\end{equation}

Similarly, pair (1)(3') gives \(\delta_{a_3a'_3}\delta_{b_3b'_3}\delta_{c_3c'_3}\psi_2^{a_3b_3c_3}\psi_2^{a'_3b'_3c'_3}\). In the large \(N\) limit, the first pair (1)(1') dominates. Multiplying by 3 - we do this because we only looked at the pairs that contain tetrahedron (1) - we find \(3N^3G(t-t')^2G(t-t_3)G(t'-t_4)\). The Greens function come from the wick contractions that we have done earlier.

Now, contracting \(\psi_1^{a_2b_2c_2}\) with \(\psi_1^{a'_2b'_2c'_2}\) and \(\psi_2^{a_3b_3c_3}\) with \(\psi_2^{a'_3b'_3c'_3}\) gives the same result, but leaves behind \(\psi_1\psi_1\). The same is true in the case in which \(\psi_1^{a_1b_1c_1}\) with \(\psi_1^{a'_1b'_1c'_1}\) and \(\psi_2^{a_3b_3c_3}\) with \(\psi_2^{a'_3b'_3c'_3}\) are contracted. This is because the diagram that one obtains after the contraction is similar to figure \ref{kernelcalc2}: The only difference is just that the face X, Y, Z are on a different face now.

\begin{figure}
\begin{center}
\includegraphics[scale=0.4]{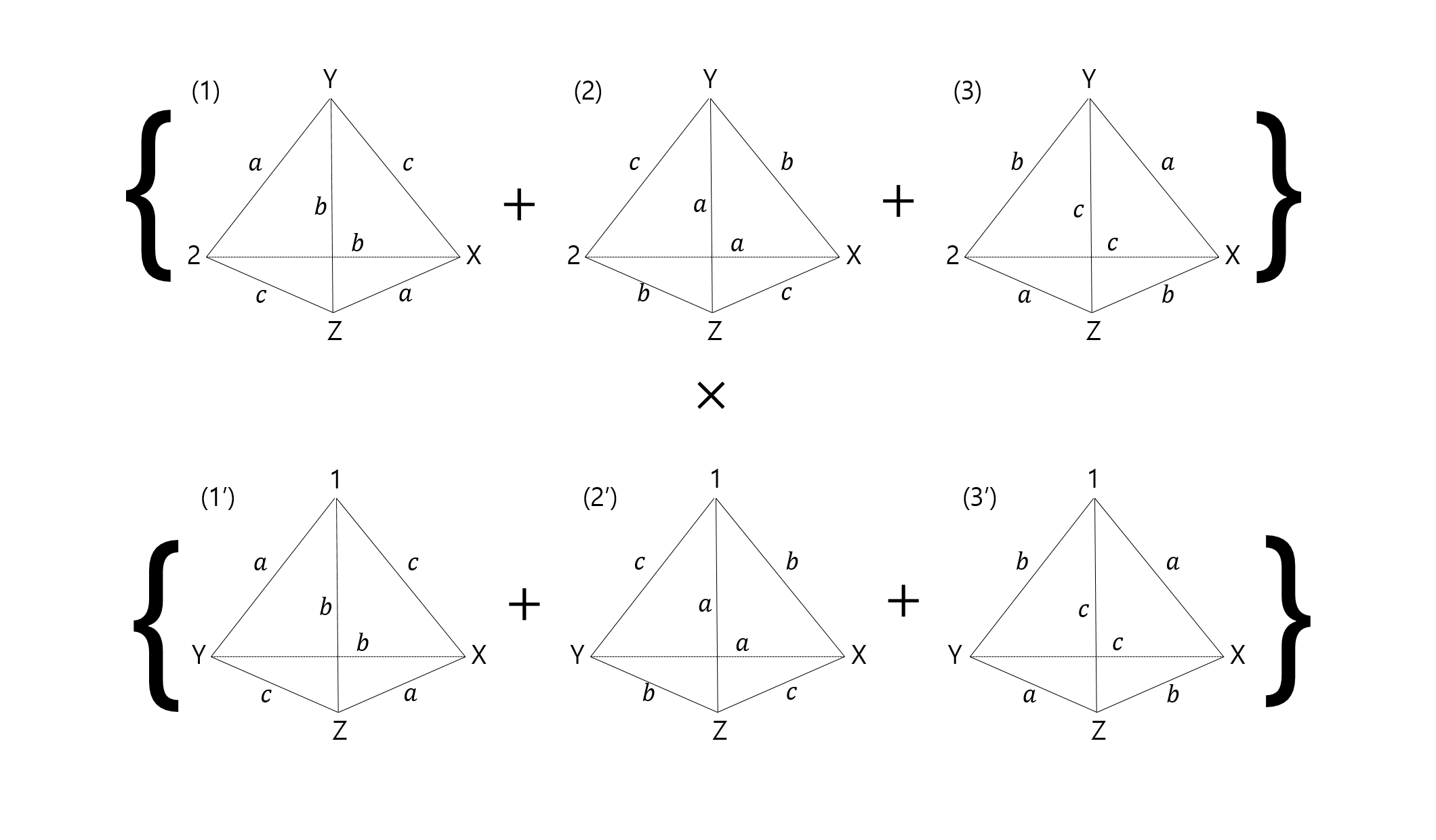}
\end{center}
\caption{The calculation of the kernel, after the contraction of \(\psi_2^{a_1b_1c_1}\) with \(\psi_2^{a'_2b'_2c'_2}\) and \(\psi_2^{a_3b_3c_3}\) with \(\psi_2^{a'_3b'_3c'_3}\).}
\label{kernelcalc3}
\end{figure}

Now, suppose we contract \(\psi_1^{a_1b_1c_1}\) with \(\psi_1^{a'_2b'_2c'_2}\) and \(\psi_2^{a_3b_3c_3}\) with \(\psi_2^{a'_3b'_3c'_3}\). This provides us with a picture as drawn in figure \ref{kernelcalc3}.

Now, try to match the face XYZ of the top row to that of the bottom row: whenever the colors \(a\), \(b\), \(c\) of the edges of the top and bottom row XYZs match with one another, it would give a factor of \(N\) for it corresponds to the contraction of two delta functions with one another. However, the maximum factor that one gets here is going to be \(N\) because there is no way to make more than two edges match one another. Hence this wick contraction gets ignored in the large \(N\) limit for it gets dwarfed by other terms.

We make a quick digression here: after doing the wick contraction, in matching the face XYZ of the top row to the bottom row, if the remaining points are on the same side, one gets a factor of \(3N^3\). On the other hand, if the remaining points are on the opposite side, one gets a factor of the order \(N\).

This is because if the the two remaining points are on the same side, then one out of the three pairs (1)(1'), (1)(2'), and (1)(3') that we look at exhibits a complete match of the colors on all sides of the faces XYZ. This gives a factor of \(N^3\). The other two pairs would be completely mismatching and hence would vanish in the large \(N\) limit. After considering pairs with tetrahedrons (2) and (3), one arrives at the factor of \(3N^3\).

On the contrary, if the two remaining points are on opposite sides, one can never arrive at a complete match of the colors on all the edges of the face XYZ of the top row to the face XYZ of the bottom row. The most that one can get is a single match, and subsequently one arrives at the factor of the order \(N\).

Consequently, the contribution of the \(I_1(a_i,b_i,c_i)I_1(a'_i,b'_i,c'_i)\) term to the kernel is

\begin{equation}
\begin{split}
\MoveEqLeft
-3\beta^2g^2N^3
\begin{bmatrix}
1 & 2 \\ 2 & 1
\end{bmatrix}
G(t_3-t)G(t_4-t')G(t-t')^2
\end{split}
\end{equation}

Where the minus sign comes from the contraction of the remaining \(\psi\) terms from the kernel with \(\psi\) terms outside of the kernel.

\newpage
\section{Conformal Field Theory}
In this section we give a quick introduction of Conformal Field Theory - giving emphasis to the two point and three point functions. What we discuss here was written after \cite{Qualls:2015qjb}.

A conformal field theory is a quantum field theory that is invariant under the conformal group. A conformal group is a group of mappings that preserve the angle. More concretely, a transformation belongs in the conformal group if the metric of the transformed coordinates is a scalar multiple of that of the original coordinates

\begin{equation}
g_{\rho \sigma}\frac{\partial x'^{\rho}}{\partial x^{\mu}}\frac{\partial x'^{\sigma}}{\partial x^{\nu}} = \Lambda(x)g_{\mu \nu}
\end{equation}

In this section we will talk about conformal invariance in flat space \(R^{d}\) where \(d \neq 2\). This is more relevant to our thesis because our theory is at \(D = 1\): With conformal groups in \(d = 2\) an infinite number of symmetries occur which are dealt with the Virasoro algebra, and this is not very useful in the objects that we deal with in our thesis.

In flat space, the Poincare group - transformations that preserve the minkowski metric - belong in the conformal group. In addition to the Poincare group, the conformal group includes scaling transformations, and the special conformal transformation

\begin{equation}
x'^{\mu} = \lambda x^{\mu}
\end{equation}
\begin{equation}
x'^{\mu} = \frac{x^{\mu} - (x \cdot x)b^{\mu}}{1-2(b \cdot x) + (b \cdot b)(x \cdot x)}
\end{equation}

It is trivial that the former is a conformal transformation, for a scaling of the coordinates \(dx'^{\mu} = \lambda dx^{\mu}\) would make the metric become \(g'_{\mu\nu} = \frac{1}{\lambda^2}g_{\mu\nu}\). The reason why the latter is conformal is a bit more complex.

Let us define the following transformation as an inversion: \(x^{\mu} = \frac{x^{\mu}}{x \cdot x}\). Then the special conformal transformation satisfies the following: \(x'^{\mu} = \frac{x'^{\mu}}{x' \cdot x'} = x^{\mu} = \frac{x^{\mu}}{x \cdot x} - b^{\mu}\). Therefore, a special conformal transformation is an inversion followed by a translation, then another inversion.

An inversion, however, is a conformal transformation: \(x^{\mu} = \frac{x^{\mu}}{x \cdot x}\), where the scalar coefficient following the transformation of the metric is \(\Lambda(x) = \frac{1}{(x \cdot x)}\). Hence, since all three components of a special conformal transformation are conformal, a special conformal transformation is a conformal transformation.

Some more differential analysis lead to the fact that the transformations that we discussed so far are all of the possible conformal transformations. Consequently, a conformal group in \(R^{p+q}\) where \(p+q \neq 2\) is composed of translations, rotations, dilatations, and special conformal transformations. Note that we do not include inversion transformations in the conformal group, since it is a discrete transformation.

An example of a conformal field theory is the massless scalar \(\phi^4\) theory at \(d = 4\). The action, given as

\begin{equation}
S = \int{d^4x} \frac{1}{2}(\partial_{\mu}\phi)^2 + \frac{\lambda}{4!}\phi^4
\end{equation}

is invariant under the conformal transformation \(x \rightarrow x'\), and \(\phi(x) \rightarrow \phi'(x')\), where

\begin{equation}
\phi'(x') = \lambda^{-\Delta}\phi(x)
\end{equation}

and \(\lambda = \left |\frac{\partial x'}{\partial x}\right |^{1/d}\), \(\Delta = 1\). \(\left |\frac{\partial x'}{\partial x}\right |\) is the jacobian of the transformation.

Note that a massive scalar \(\phi^4\) theory is not conformal, for the same transformation is going to make the integral of the mass term be of \(\lambda^2\).

The scaling dimension of the operator \(\phi\) is \(\Delta = 1\). In general, the scaling dimension of an operator \(O\) is defined as the action of a dilatation on the operator \(O\)

\begin{equation}
O(\lambda x) = \lambda^{-\Delta}O(x) 
\end{equation}

Now conformal invariance in CFTs impose some important restrictions on the two-point functions and the three-point functions. Let us first look at two-point functions

\begin{equation}
\begin{split}
\MoveEqLeft
\braket{\phi_1(x_1)\phi_2(x_2)} = \frac{1}{Z}\int{D\Phi}\phi_1(x_1)\phi_2(x_2)e^{-S[\Phi]} \\& \\& \qquad
= \left |\frac{\partial x'}{\partial x}\right |^{\Delta_1/d}_{x=x_1} \left |\frac{\partial x'}{\partial x}\right |^{\Delta_2/d}_{x=x_2} \frac{1}{Z}\int{D\Phi}\phi_1(x'_1)\phi_2(x'_2)e^{-S[\Phi]} \\& \\& \qquad
= \left |\frac{\partial x'}{\partial x}\right |^{\Delta_1/d}_{x=x_1} \left |\frac{\partial x'}{\partial x}\right |^{\Delta_2/d}_{x=x_2}\braket{\phi_1(x_1)\phi_2(x_2)}
\label{cfttwo}
\end{split}
\end{equation}

Now, rotation invariance implies \(\braket{\phi_1(x_1)\phi_2(x_2)} = f(|x_1-x_2|)\). Furthermore, invariance to dilatations \(x \rightarrow \lambda x\) indicate that \(f(\lambda x)=\lambda^{-(\Delta_1+\Delta_2)}f(x)\). Therefore, \(\braket{\phi_1(x_1)\phi_2(x_2)} = \frac{d_{12}}{|x_1-x_2|^{\Delta_1 +\Delta_2}}\)

Doing a special conformal transformation, we find that only when \(\Delta_1 = \Delta_2\) can equation \ref{cfttwo} can hold, and consequently

\begin{equation}
\braket{\phi_1(x_1)\phi_2(x_2)} = \frac{d_{12}\delta_{\Delta_1 \Delta_2}}{|x_1-x_2|^{\Delta_1+\Delta_2}}
\end{equation}

Now let us look at three point functions. Similar to what was done above, due to invariance under translations, rotations, dilatations, and special conformal transformations, the three point function is forced to satisfy the following equation

\begin{equation}
\braket{\phi_1(x_1)\phi_2(x_2)\phi_3(x_3)} = \frac{\lambda_{123}}{|x_{12}|^{\Delta-2\Delta_3}|x_{23}|^{\Delta-2\Delta_1}|x_{31}|^{\Delta-2\Delta_2}}
\end{equation}

Where \(x_{ij} = |x_i - x_j|\), and \(\Delta = \Delta_1+\Delta_2+\Delta_3\). These \(\lambda_{123}\) is important in defining a CFT. We quickly sketch the reason why.

In a conformal field theory, a conformal primary operator is an operator which commutes with \(K_{\mu}\). Now there is a correspondence between an operator and a state. This correspondence leads to the operator product expansion, in which means that a product of two operators can be expressed as a sum of conformal primaries with some coefficients: these coefficients are unique, and hence defines a CFT. \cite{Qualls:2015qjb}

\section{Useful Integrals}
The following integral comes in handy in calculating the spectra of the ladder diagrams \cite{Klebanov:2016xxf}

\begin{equation}
\begin{split}
\MoveEqLeft
\int_{-\infty}^{\infty}{du \frac{\sgn(u-t_1)\sgn(u-t_2)}{|u-t_1|^a|u-t_2|^b}} = l_{a,b}^+\frac{1}{|t_1-t_2|^{a+b-1}}
\\&
\int_{-\infty}^{\infty}{du \frac{\sgn(u-t_2)}{|u-t_1|^a|u-t_2|^b}} = l_{a,b}^-\frac{\sgn(t_1-t_2)}{|t_1-t_2|^{a+b-1}}\
\end{split}
\end{equation}
Where, \(l_{a,b}^{\pm} = \beta (1-a, a+b-1) \pm \big(\beta (1-b,a+b-1) - \beta (1-a, 1-b)\big)\)

For the operators $O_{1,2} = \psi_1 \partial_t^{2n+1} \psi_1 \pm \psi_2 \partial_t^{2n} \psi_2$ and $O_4 = \psi_1 \partial_t^{2n} \psi_2 - \psi_2 \partial_t^{2n} \psi_1$, the following integral comes in handy

\begin{equation}
\begin{split}
\MoveEqLeft
\int{dtdt'}\frac{\sgn(t-t')}{|t-t'|^{1/2-h}}\frac{\sgn(t-t_1)\sgn(t'-t_2)}{|t-t_1|^{1/2}|t'-t_2|^{1/2}|t-t'|}\\&\qquad
= \int{dtdt'}\frac{\
\sgn(t-t')\sgn(t-t_1)\sgn(t-t_2)}{|t-t_1|^{1/2}|t'-t_2|^{1/2}|t-t'|^{3/2-h}} \\& \qquad
= l^+_{\frac{3}{2}-h,\frac{1}{2}}l^-_{1-h, \frac{1}{2}}\frac{\sgn(t_1-t_2)}{|t_1-t_2|^{1/2 - h}} \\& \qquad
= 2\pi \frac{\tan(\frac{\pi}{2}(h-\frac{1}{2}))}{h-1/2}\frac{\sgn(t_1-t_2)}{|t_1-t_2|^{1/2 - h}}
\end{split}
\end{equation}

For the operator $O_3 = \psi_1 \partial_t^{2n} \psi_2 + \psi_2 \partial_t^{2n} \psi_1$, this integral comes in handy

\begin{equation}
\begin{split}
\MoveEqLeft
\int{dtdt'}\frac{1}{|t-t'|^{1/2-h}}\frac{\sgn(t-t_1)\sgn(t'-t_2)}{|t-t_1|^{1/2}|t'-t_2|^{1/2}|t-t'|} \\&\qquad = \int{dtdt'}\frac{\sgn(t-t_1)\sgn(t-t_2)}{|t-t_1|^{1/2}|t'-t_2|^{1/2}|t-t'|^{3/2-h}} \\& \qquad
= l^-_{\frac{3}{2}-h,\frac{1}{2}}l^+_{\frac{1}{2},1-h}\frac{1}{|t-t'|^{1/2-h}} \\& \qquad
= 2\pi\frac{\tan(\frac{\pi}{2}(h+\frac{1}{2}))}{h-\frac{1}{2}}\frac{1}{|t-t'|^{1/2-h}} \\& \qquad
\end{split}
\end{equation}

\newpage

\section*{Honor Code and Authorizations}
\begin{center}
This senior thesis represents my own work in accordance with University Regulations.

I authorize Princeton University to lend this thesis to other institutions or individuals for the purpose of scholarly research.

I further authorize Princeton Univeristy to reproduce this thesis by photocopying or other means, in total or in part, at the request of other institutions or individuals for the purpose of scholarly research.

\vspace{5mm}

Jaewon Kim
\end{center}

\section*{Acknowledgements}	
First and foremost, I would like to thank my advisor Igor Klebanov, without whose guidance this thesis would have been impossible. Over the course of my junior spring and senior year, his teachings on quantum field theory and large N tensor models has made me love physics: His passion for physics is inspiring, and I aspire to become a physicist as passionate as he. I thank him for his willingness to participate in discussions, for his helpful and speedy feedbacks, and for the opportunities he has given me. I cannot express in words how much working with him has meant to me.

I also owe the smoothness of this thesis to Grigory Tarnopolsky, who minutely reviewed my rough draft. His comments were invaluable in the shaping of this thesis.

The discussions with Prof. Silviu Pufu, Vladimir Kirilin, and Ksenia Bulycheva have also been invaluable to me. Being the spontaneous person that I am, whenever I would have questions I would peak around the corner of their offices; they always welcomed me, and patiently held discussions with me until they were sure that I had understood the answers to my questions. Thank you so much - I owe my understanding of Conformal Field Theory, and the SYK model to you.

I would also like to express my gratitude towards Prof. Shivaji Sondhi, and Prof. Simone Giombi. My junior independent works with them were my greatest sources of fun during my junior year. Without their guidance and love, I would not be where I am right now. Their teachings have nurtured me into a physicist, and I will be forever indebted to them.

On a broader scale, I would like to thank the department of physics. I will cherish the connections that I have made with the faculty, staff, and students over the course of the past four years as I start the next step of my career at Berkeley. I am very grateful for all that they have done for me.

The Korean Student Aid Foundation funded most of my education at Princeton University, and I would like to take a moment here to thank them for their generosity. Without their help, I would not have been able to attend this wonderful University.

I am also indebted to all my friends; they have made my life at Princeton so wonderful. Brandon, you were the greatest roommate that I could have ever wished for. Josh, thank you for accompanying me in so many runs and climbing sessions. Also, I owe much thanks to Helen, Jot, Josh, Andrew, Louis, Jeff, Kyle, Julie, Ben, and friends from the Brown Coop, Princeton Climbing Team, Mathey, and Quadrangle Club. They have given me so much joy throughout my four years at Princeton, and were my family away from my family.

I want to next thank Maggie Pecsok, my amazing partner. You are the best thing that has happened to me at Princeton: your love have transformed my life for the better, and I am excited to begin my life outside of the orange bubble with you. I love you, Maggie.

Lastly, I must thank my sister, Siyoon, and my parents. I could never repay what they have gave and sacrificed for me. Their wisdom and love have propelled me throughout my years at Princeton, and will continue to do so afterwards. Thank you for everything Mum, dad, and sis. I love y'all so so much.

Yours Sincerely,

\vspace{10mm}

Jaewon Kim

\newpage

\bibliography{citations}

\begin{thebibliography}{10}

\bibitem{Ambjorn:1992iz}
Jan Ambjorn, C.~F. Kristjansen, Z.~Burda, and J.~Jurkiewicz.
\newblock {Three-dimensional simplicial quantum gravity coupled to Ising
  matter}.
\newblock {\em Nucl. Phys. Proc. Suppl.}, 30:771--774, 1993.

\bibitem{Bonzom:2011zz}
Valentin Bonzom, Razvan Gurau, Aldo Riello, and Vincent Rivasseau.
\newblock {Critical behavior of colored tensor models in the large N limit}.
\newblock {\em Nucl. Phys.}, B853:174--195, 2011, 1105.3122.

\bibitem{Brezin:1977sv}
E.~Brezin, C.~Itzykson, G.~Parisi, and J.~B. Zuber.
\newblock {Planar Diagrams}.
\newblock {\em Commun. Math. Phys.}, 59:35, 1978.

\bibitem{Bulycheva:2017ilt}
Ksenia Bulycheva, Igor~R. Klebanov, Alexey Milekhin, and Grigory Tarnopolsky.
\newblock {Spectra of Operators in Large $N$ Tensor Models}.
\newblock {\em Phys. Rev.}, D97(2):026016, 2018, 1707.09347.

\bibitem{Carrozza:2015adg}
Sylvain Carrozza and Adrian Tanasa.
\newblock {$O(N)$ Random Tensor Models}.
\newblock {\em Lett. Math. Phys.}, 106(11):1531--1559, 2016, 1512.06718.

\bibitem{Gross:2016kjj}
David~J. Gross and Vladimir Rosenhaus.
\newblock {A Generalization of Sachdev-Ye-Kitaev}.
\newblock {\em JHEP}, 02:093, 2017, 1610.01569.

\bibitem{Gross:2017aos}
David~J. Gross and Vladimir Rosenhaus.
\newblock {All point correlation functions in SYK}.
\newblock {\em JHEP}, 12:148, 2017, 1710.08113.

\bibitem{Gross:1991hx}
Mark Gross.
\newblock {Tensor models and simplicial quantum gravity in > 2-D}.
\newblock {\em Nucl. Phys. Proc. Suppl.}, 25A:144--149, 1992.

\bibitem{Gubser:1998bc}
S.~S. Gubser, Igor~R. Klebanov, and Alexander~M. Polyakov.
\newblock {Gauge theory correlators from noncritical string theory}.
\newblock {\em Phys. Lett.}, B428:105--114, 1998, hep-th/9802109.

\bibitem{Gurau:2009tw}
Razvan Gurau.
\newblock {Colored Group Field Theory}.
\newblock {\em Commun. Math. Phys.}, 304:69--93, 2011, 0907.2582.

\bibitem{Gurau:2016lzk}
Razvan Gurau.
\newblock {The complete $1/N$ expansion of a SYK–like tensor model}.
\newblock {\em Nucl. Phys.}, B916:386--401, 2017, 1611.04032.

\bibitem{Gurau:2011aq}
Razvan Gurau and Vincent Rivasseau.
\newblock {The 1/N expansion of colored tensor models in arbitrary dimension}.
\newblock {\em EPL}, 95(5):50004, 2011, 1101.4182.

\bibitem{Gurau:2011xp}
Razvan Gurau and James~P. Ryan.
\newblock {Colored Tensor Models - a review}.
\newblock {\em SIGMA}, 8:020, 2012, 1109.4812.

\bibitem{Kitaev:2015}
Alexei Kitaev.
\newblock {A simple model of quantum holography}.
\newblock
  \url{http://online.kitp.ucsb.edu/online/entangled15/kitaev/},\url{http://online.kitp.ucsb.edu/online/entangled15/kitaev2/}.
  Talks at KITP, April 7, 2015 and May 27, 2015.

\bibitem{Klebanov:2018fzb}
Igor~R. Klebanov, Fedor Popov, and Grigory Tarnopolsky.
\newblock {TASI Lectures on Large $N$ Tensor Models}.
\newblock {\em PoS}, TASI2017:004, 2018, 1808.09434.

\bibitem{Klebanov:2016xxf}
Igor~R. Klebanov and Grigory Tarnopolsky.
\newblock {Uncolored random tensors, melon diagrams, and the Sachdev-Ye-Kitaev
  models}.
\newblock {\em Phys. Rev.}, D95(4):046004, 2017, 1611.08915.

\bibitem{Maldacena:2016hyu}
Juan Maldacena and Douglas Stanford.
\newblock {Comments on the Sachdev-Ye-Kitaev model}.
\newblock {\em Phys. Rev.}, D94(10):106002, 2016, 1604.07818.

\bibitem{Maldacena:1997re}
Juan~Martin Maldacena.
\newblock {The Large N limit of superconformal field theories and
  supergravity}.
\newblock {\em Int. J. Theor. Phys.}, 38:1113--1133, 1999, hep-th/9711200.
\newblock [Adv. Theor. Math. Phys.2,231(1998)].

\bibitem{Moshe:2003xn}
Moshe Moshe and Jean Zinn-Justin.
\newblock {Quantum field theory in the large N limit: A Review}.
\newblock {\em Phys. Rept.}, 385:69--228, 2003, hep-th/0306133.

\bibitem{Polchinski:2016xgd}
Joseph Polchinski and Vladimir Rosenhaus.
\newblock {The Spectrum in the Sachdev-Ye-Kitaev Model}.
\newblock {\em JHEP}, 04:001, 2016, 1601.06768.

\bibitem{Qualls:2015qjb}
Joshua~D. Qualls.
\newblock {Lectures on Conformal Field Theory}.
\newblock 2015, 1511.04074.

\bibitem{Sachdev:1992fk}
Subir Sachdev and Jinwu Ye.
\newblock {Gapless spin fluid ground state in a random, quantum Heisenberg
  magnet}.
\newblock {\em Phys. Rev. Lett.}, 70:3339, 1993, cond-mat/9212030.

\bibitem{Sasakura:1990fs}
Naoki Sasakura.
\newblock {Tensor model for gravity and orientability of manifold}.
\newblock {\em Mod. Phys. Lett.}, A6:2613--2624, 1991.

\bibitem{'tHooft:1973jz}
Gerard 't~Hooft.
\newblock {A Planar Diagram Theory for Strong Interactions}.
\newblock {\em Nucl. Phys.}, B72:461, 1974.

\bibitem{Witten:1998qj}
Edward Witten.
\newblock {Anti-de Sitter space and holography}.
\newblock {\em Adv. Theor. Math. Phys.}, 2:253--291, 1998, hep-th/9802150.

\bibitem{Witten:2016iux}
Edward Witten.
\newblock {An SYK-Like Model Without Disorder}.
\newblock 2016, 1610.09758.

\end{thebibliography}
\bibliographystyle{hplain}

\end{document}